\documentclass[aps,prd,noshowpacs,preprintnumbers,nofootinbib,floatfix,twoside]{revtex4}

\usepackage{amsmath}
\usepackage{amssymb}
\usepackage{graphicx}
\usepackage{bm}
\usepackage{subfigure}

\newcommand{\Cal}[1]{\ensuremath{\mathcal{#1}}}

\newcommand{\eqn}[1]{Eqn. \eqref{#1}}
\newcommand{\eqns}[1]{Eqns. \eqref{#1}}
\newcommand{\Cite}[1]{Ref. \cite{#1}}
\newcommand{\Cites}[1]{Refs. \cite{#1}}
\newcommand{\Sec}[1]{Sec. \ref{#1}}
\newcommand{\fig}[1]{Fig. \ref{#1}}

\newcommand{\tab}[1]{Table \ref{#1}}

\newcommand{\ti}[1]{{\tilde #1}}

\newcommand{\be}{\begin{equation}}
\newcommand{\ee}{\end{equation}}
\newcommand{\la}[1]{\label{#1}}

\newcommand{\p}{\ensuremath{\partial}}

\newcommand{\eps}{\ensuremath{\epsilon}}
\newcommand{\om}{\ensuremath{\omega}}
\newcommand{\lam}{\ensuremath{\lambda}}
\newcommand{\state}[1]{\ensuremath{|{\tt #1}\rangle}}
\newcommand{\vphi}{\ensuremath{\varphi}}
\newcommand{\vev}[2]{\ensuremath{\langle{\tt #1}| #2 |{\tt
      #1}\rangle}} 
\newcommand{\avg}[1]{\ensuremath{\langle\,#1\,\rangle}}

\begin{document}

\title{Radiation from collapsing shells, semiclassical backreaction
  and black hole formation} 

\author{Aseem Paranjape}
\email{aseem@tifr.res.in}
\affiliation{Tata Institute of Fundamental Research, Homi Bhabha
Road, Colaba, Mumbai - 400 005, India\\}

\author{T. Padmanabhan}
\email{paddy@iucaa.ernet.in}
\affiliation{IUCAA, Post Bag 4, Ganeshkhind, Pune - 411 007, India\\}

\date{\today}

\begin{abstract}
\noindent
We provide a detailed analysis of quantum field theory around a
collapsing shell and discuss several conceptual issues related to the
emission of  radiation flux  and formation of black holes. Explicit
calculations are performed using a model for a collapsing shell which
turns out to be analytically solvable. We use the insights gained in
this model to draw reliable conclusions regarding more realistic
models. We first show that any shell of mass $M$ which collapses to a
radius close to $r=2M$ will emit approximately thermal radiation for a
period of  time. In particular, a shell which collapses from some
initial radius to a final radius $2M(1-\epsilon^2)^{-1}$ (where
$\epsilon \ll 1$) without forming a black hole, will emit thermal
radiation during the period $M\lesssim t \lesssim M\ln
(1/\epsilon^2)$. Later on ($t\gg M \ln(1/\epsilon^2)$), the flux from
such a shell will decay to zero exponentially.  We next study the
effect of backreaction computed using the vacuum expectation value of
the stress tensor on the collapse.  We find that, in any realistic
collapse scenario, the backreaction effects do \textit{not} prevent
the formation of the event horizon. The time at which the event
horizon is formed is, of course, delayed due to the radiated flux ---
which decreases the mass of the shell --- but this effect is not
sufficient to prevent horizon formation. We also
clarify several conceptual issues and provide pedagogical details of
the calculations in the Appendices to the paper. 
\end{abstract}

\maketitle

\section{Introduction, Motivation and Summary}
\noindent
Classical general relativity allows for solutions in which matter can
collapse in a spherically symmetric manner, forming a black hole event
horizon when viewed from the outside region. The collapsing matter
hits a spacetime singularity in finite proper time thereby preventing
us from using the classical equations to predict the future evolution
of the system, as viewed by an observer collapsing with the matter. On
the other hand, as far as any outside observer is concerned, the
collapsing material takes infinite time (as measured by the stationary
clocks outside) to reach the event horizon and hence the formation of
the singularity  has no influence on the outside region. While this
may appear bizarre to the uninitiated, general relativists have
learned to live with this dichotomous evolutionary behaviour of the
system.  

A new layer of complication arises
when we go beyond the classical theory and study quantum fields
evolving in the background geometry of collapsing matter.  If a
quantum field is in the vacuum state in the asymptotic past, then at
late times an observer at spatial infinity will see a flux of energy
from the collapsing matter corresponding to a thermal radiation with
temperature $T_H=1/8\pi M$ \cite{hawking}. (Throughout the paper,  we
will use Planckian units with $G$, $\hbar$ and $c$ set to 
unity.) The thermal nature is closely related to
the exponential redshift experienced by the modes when they travel
from a region close to $r=2M$ to $r\to\infty$ and --- to that extent
--- the thermal nature of the radiation depends on  matter collapsing
to a size $r\approx 2M$. This process has been  extensively
investigated and is well understood as long as we treat the quatum
field as a test field (see \Cite{visser-bh-ess} for a pedagogical
treatment and a review of the literature). 

But the radiation flux to $r\to\infty$ is `real' in the sense that the
observer can collect it and use it to do work (say, e.g., to heat up
some water). Energy conservation then requires that the mass of the
collapsing body  must decrease due to the outgoing radiation, since it
is the only possible source of energy. In that case, we cannot treat
the background as fixed but must solve for Einstein's equations with a
semiclassical backreaction term added (which is usually modeled
through the vacuum expectation value (VEV) of the stress tensor with
the adjective `vacuum' referring to the choice that, in the asymptotic
past, the quantum field was in the vacuum state) and solve for the
radiated flux simultaneously and self-consistently. Several authors
have studied this phenomenon
\cite{brout,bir-dav,page,sanchez,howard,davies,bardeen,york,unruh,
  massar,brout-appD,clifton,brown,nielsen}  
and  in particular we refer to the review 
by Brout et al. \cite{brout}. As long as the flux of particles,
characterized by the VEV of the stress tensor  is \emph{small} and
\emph{slowly varying} (in Planckian units), the backreaction can be
self-consistently accounted for and studied in the geometry of a black
hole with a \emph{slowly varying mass}. In such a case, the
backreaction effects are not significant and can be incorporated in a
perturbative matter.  

Recently there have been claims in the literature \cite{vachaspati}
that the number of particles reaching an observer at large distances
from the black hole, at late times, can diverge. It has then been
conjectured that the backreaction from this diverging number of
particles might be sufficient to prevent the formation of the event
horizon. These claims are clearly contradictory to the usually
accepted view mentioned above, and other authors have also questioned
this view (see, e.g. \Cites{visser-liberati,visser-bh-status}; see
also \Cite{alberghi}). In the
light of such claims, we wish to revisit this issue in this paper. 

Right at the outset, let us clarify one elementary --- but potentially
misleading --- aspect of the problem. During the classical
gravitational collapse, although the collapsing object will cross its
Schwarzschild radius (and thus form an event horizon) in finite proper
time, an observer at rest at large distances from the collapsing
object will see the collapsing body approach its Schwarzschild radius
only asymptotically as $R(t)=2M(1+\Cal{O}(e^{-t/4M}))$ where $t$ is
the usual Schwarzschild time coordinate of the outside metric 
\be
ds^2_{\rm Schw} = -(1-2M/r)dt^2 + \frac{dr^2}{(1-2M/r)} +
r^2d\Omega^2\,. 
\la{1eq1}
\ee
(Throughout the paper,  we shall
deal with situations in which a spherical body collapses without
rotation.) In the absence of any backreaction, a calculation of the
stress tensor VEV will show a steady flux at late times corresponding
to a (nearly) thermal spectrum with a temperature $T_H\propto1/M$. If
one insists on waiting for an arbitrarily long time \emph{without
  accounting for backreaction}, we will clearly end up with an
arbitrarily large number of particles reaching large distances at late
enough times. Such a divergence, of course, has nothing to do with
black holes, and will arise for \emph{any} luminous object if we
pretend that its mass does not reduce due to the energy it
radiates. For example, if the sun could radiate a steady flux of
photons for an infinite amount of time without changing its mass or 
constitution, it would emit an infinite amount of energy! That is,
this situation will arise regardless of whether the event horizon
forms or not and the divergence merely indicates the need to correctly
account for the backreaction (or, more simply, energy conservation).

In principle, it is therefore inconsistent to assume that the exterior 
geometry is described by an exact Schwarzschild metric for two
reasons: First, the mass of the central object is changing and second,
the presence of an outgoing flux even at large distances means that
the geometry is not strictly asymptotically flat. One could now
imagine a situation like the following : the mass loss of the
collapsing object (in our case, the shell) implies that the radius to
which the shell must collapse to form an event horizon, is
shrinking. This means that the shell spends a larger amount of time
outside the event horizon, all the while radiating particles to
infinity. All this is likely to be further complicated by the fact 
that if the mass loss rate is high enough, the exterior geometry is
likely to be very complicated. The radius which the shell is chasing
keeps reducing, and the possibility arises that this runaway process
ends with the shell completely evaporating \emph{before} the event
horizon is formed. If this situation is generic, it would mean that
black holes typically \emph{do not} form, contrary to the popular
view. It is therefore worth investigating this scenario to settle the
effect of backreaction, and in this paper we will do this using some
simple models for collapse. The simplifications we introduce to make
the problem tractable are the following:  

(a) We assume that the collapsing system is a thin spherical shell
with internal stresses arranged in such a manner as to allow it to
collapse along some given trajectory in the spacetime. Then the
internal metric is Minkowski and the outside is Schwarszchild with
matching conditions determining either metric in terms of the other
and the trajectory of the shell \cite{unruh}.  

(b) We will assume that all computations can be performed in the $1+1$
sector of the spacetime which ignores the two angular
coordinates. This corresponds to using only the s-wave mode of the
scalar field  and allows one to use the tools of conformal field
theory to compute the VEV of the stress tensor. This assumption is
unlikely to influence the conclusions of the paper and is justified in
\Cite{brout}. 

(c) We will assume that the semiclassical backreaction
can be modeled by the VEV of the stress tensor, $<T_{ab}>$ 
with Einstein's equations modified to
$G_{ab}=8\pi(T_{ab}+<T_{ab}>)$. There is general agreement that such
an equation should be valid in some suitable limit (though no one has
rigorously proved it; but see \Cite{hartletp}) and we will proceed
hoping for the best. 

Given these assumptions, the problem can in principle be reduced to
one of solving a set of equations. Given any trajectory of the shell,
one can compute the VEV of $T_{ab}$ everywhere using the conformal
field theory technique. The problem then reduces to calculating (a)
the backreaction on the shell trajectory and (b) the flux radiated to
infinity. In practice, unfortunately, the equations turn out to be
quite intractable and one needs to obtain insights into what is
happening using simplified models. We do this along the following
lines in this paper. 

We first consider the role of the event horizon in the emission of
particles to large distances (ignoring the effect of the
backreaction). We conclude that operationally, its role depends on
when the observer at infinity sets out to detect the particles. To
understand this result, consider two collapse scenarios (say, A and B)
in which shells (each of mass $M$) start from the same initial
radius $R_0$ with further evolution being different: (a) In Case A, the
shell continuously collapses and forms an event horizon. (The shell
crosses $r=2M$ at a finite proper time as shown by the clock on the
surface of the shell but after infinite amount of coordinate time as
measured by the clocks of the observer outside the shell.) Standard
calculations indicate that at late times the asymptotic observer will
see a thermal flux of radiation.  (b) In Case B, the shell
follows exactly the same trajectory as in Case A until it reaches
close to $r=2M$. Its trajectory then deviates from that of Case A; the
shell progressively slows down, and asymptotically approaches a final
radius $r=2M (1-\epsilon^2)^{-1}$. Obviously no event horizon is
formed in this Case B  for any value of $\epsilon$, however small. It
is also clear that at very late times the geometry is static in Case B
with the shell staying at a fixed radius outside the event horizon and
there should be no flux of radiation.  But for an arbitrarily large
period of time we can arrange matters such that the trajectory in Case
B behaves similar to that in Case A when it is hovering just outside 
$r=2M$ (as seen by an outside observer). The outgoing modes which
cross the shell and propagate to future null infinity will lead to a
thermal flux in this case as well. In fact, this is an elementary
consequence of causality. What an observer at infinity sees at some
given event $\mathcal{P}$ (at time $t_{obs}$) can only depend on the
behaviour of shell trajectory which is contained within the backward
light cone of $\mathcal{P}$. The future trajectory of the shell --- in
particular whether it settles down at $r\gtrsim 2M$ or goes on to
collapse through $r=2M$ --- should not affect observations at
$\mathcal{P}$. 

We therefore expect the Case B to be characterized by three
distinct phases. In the first phase, the shell collapses from a large
radius to some radius close to $2M$ but larger than its final
asymptotic radius. In this phase, the trajectory of the
shell is identical to that of Case A. We, therefore, expect a flux of
radiation being emitted to infinity which starts from zero and builds
up to the thermal flux value characterized by the temperature $T_H
\propto (1/M)$ in the timescale of the collapse (which is
$\mathcal{O}(M)$ for nearly geodesic motion). In the second phase, the
shell hovers close to $r=2M$ \emph{before} settling down to its
asymptotic radius. If we choose the trajectory such that the timescale
governing this phase is $\mathcal{O} (M)$, then during this phase also
the trajectory is similar to that of Case A, and now we would expect
an approximately thermal flux being emitted by the shell just as
though it is going to collapse to $r=2M$. The crucial difference
between the two cases arise in the third phase, during which the shell
is asymptotically coming to rest. \textit{We will show by explicit 
calculation and  a numerical analysis that during this phase the flux
of radiation from the body dies down to zero exponentially.} If the
first two phases are characterized by a timescale $\mathcal{O} (M)$,
then the intermediate phase during which the shell emits approximately
thermal radiation lasts for a timescale of the order of $M \ln
(1/\epsilon^2)$.  

This explicit demonstration is one of the key results of this paper.
This result shows that the thermal flux of radiation observed from a
collapsing structure is not directly dependent on the formation of the
event horizon. When the event horizon does form, the thermal radiation
continues to escape for infinite amount of time  if the backreaction
effects are ignored. This is, of course, unrealistic since it would
require an infinite source of energy. So in any realistic collapse
scenario, the emission of thermal radiation is an idealization which
ceases to be valid at sufficiently late times. Our model calculation
shows that a similar effect can be mimicked by having a system
collapse to a radius $r=2M(1-\epsilon^2)^{-1}$ asymptotically. The
approximately thermal radiation will emanate from the body during the
time interval $M\lesssim t \lesssim M\ln (1/\epsilon^2)$. At very late
times ($t\gg M\ln (1/\epsilon^2)$) the radiation dies down
exponentially. 

The second key result of the paper is related to the computation of
semiclassical backreaction and its effect on the formation of event
horizon. We show that the effect of backreaction is essentially to
delay the formation of event horizon by an amount $\delta v$ (where
$v$ is the in-going Eddington-Finkelstein null coordinate) given by  
\be
\delta v = \Cal{O}\left(
\frac{M}{\beta_h^2}L_H\right) \,,
\la{3eq11-intro}
\ee
where $\beta_h$ is the proper velocity with which
the unperturbed trajectory crosses $r=2M$ and $L_H$ is the luminosity 
of the thermal radiation. This result is intuitively
clear : The delay in the event horizon formation is governed by the
product of $L_H$ and the timescale of the unperturbed collapse, which
is essentially set by the initial mass of the collapsing
object. Physically this is simply the naive bound one might place on
the amount of mass that the collapsing object can radiate via
semiclassical emission, during the collapse. The velocity at
(unperturbed) horizon crossing $\beta_h$ is completely determined by
the initial conditions (namely that the shell starts collapsing with
zero velocity at a radius $r=R_0$ at initial time). For example, for 
geodesic infall starting at $r=R_0$, one has $\beta_h^2=1-2M/R_0$, and
typically one expects $\beta_h\lesssim\Cal{O}(1)$ for more general
trajectories as well. For all such trajectories the backreaction,
which varies on a timescale $\sim M^3$, \emph{cannot prevent} the
formation of the event horizon which occurs on a timescale $\sim M$;
at best it can delay this formation by a time $\sim1/M$.
(One may wonder whether one can arrange matters for $\beta_h$ to
become sufficiently small to increase this timescale arbitrarily. This
looks implausible and we provide detailed arguments in the relevant
section eliminating this possibility.) 

It is possible to use these
results to also put a bound on the total amount of mass that can be
radiated away before the event horizon is formed and we find that it
is given by $M^{-1}\ln M$ for most of the realistic
trajectories. These results strongly suggest that the conventional
wisdom as regards the formation of black hole is correct and that the 
semiclassical radiation does not prevent the formation of the event
horizon.

The plan of the paper is as follows. Section \ref{shellradn} deals
with the computation of the renormalized stress tensor VEV for a
collapsing shell, in the \emph{absence} of backreaction. We first
recall some basic concepts of quantum fields in curved spaces in
\Sec{hawkingbasics}. To set the stage, in \Sec{null} we analyse a
trajectory which does not form a horizon and whose collapsing phase is 
null. While being unrealistic, this toy example admits of an exact
calculation of the stress tensor VEV. In \Sec{timelike} we turn to the 
more realistic timelike trajectory discussed above, and analyse the
behaviour of the stress tensor VEV in various phases of the collapse
(which, again, does not form a horizon).

In section \ref{bkrxn} we include the effects of the backreaction of
the stress tensor VEV, on the background geometry. We begin by
recalling details of a self-consistent picture as presented by Brout
et al. \cite{brout}, which describes the exterior geometry in the
presence of a \emph{small} and \emph{slowly varying} emission from the
collapsing object. We then present a simple calculation based on this
picture, to derive the expression \eqref{3eq11-intro} for the delay in
formation of the event horizon in the presence of backreaction. We
argue that even for (classical) trajectories which slow down
drastically as they approach the unperturbed Schwarzschild radius, it
is highly implausible that backreaction can significantly delay the
formation of the event horizon. Finally, we calculate an upper bound
on the mass that can be radiated away due to semiclassical emission,
using the extreme trajectories studied in \Sec{shellradn}. We briefly
highlight our main results and conclude in section
\ref{conclude}. Appendix A contains pedagogical details of results
from 2-dimensional conformal field theory which are relevant to our
calculations, and Appendix B contains detailed proofs of various
results that are quoted in the main text.

\section{Flux of radiation from a collapsing shell}
\label{shellradn}
\noindent
We begin by recalling several results related to the quantization of a  
scalar field   propagating on a fixed
background geometry, which is taken to be that of a collapsing
object. At the zeroth order we treat the scalar field as a test field
so that the stress-energy tensor of the scalar 
field does not  affect the background.  For
simplicity, throughout the paper we will consider the collapse of a
thin shell of mass $M$, so that the exterior geometry is described by
the Schwarzschild metric, while the interior is flat Minkowski
spacetime. Since we will eventually be mainly interested in order of
magnitude estimates and asymptotic behaviour of various quantities, we
will focus on broad properties of the collapse trajectory without
going into details of the shell stress-energy tensor, etc. 

\subsection{Background geometry and mode analysis}
\label{hawkingbasics}
\noindent
The exterior geometry of the shell is given by the Schwarzschild
metric, which we write using the Eddington-Finkelstein coordinates, as 
\begin{align}
ds^2_{(\rm ext)} &= -(1-\frac{2M}{r})dudv + r^2(u,v)d\Omega^2
\nonumber\\ 
& = -(1-2M/r)dv^2 + 2dvdr +r^2d\Omega^2\,,
\la{2eq1}
\end{align}
where $M$ is the (constant) mass of the shell and $r$ is defined
implicitly via 
\be
e^{(v-u)/4M} = \left(\frac{r}{2M}-1\right) e^{r/2M}~~;~~ r>R_s\,, 
\la{2eq2}
\ee
where $r=R_s(\tau)$ is the trajectory of the shell with proper time
$\tau$. The interior geometry is that of Minkowski spacetime, whose
metric we write as  
\be
ds^2_{(\rm int)} = -dT^2 + dr^2 + r^2d\Omega^2 = -dUdV +
r^2(U,V)d\Omega^2 = -dV^2 + 2dVdr + r^2d\Omega^2\,,
\la{2eq3}
\ee
where we have defined the light cone coordinates $V=T+r$, $U=T-r$, and
we use the same $r$ coordinate to label 2-spheres in the interior and
exterior. These metrics can be matched at the surface of the shell to
get equations for $U(u)$ and $V(v)$ which are valid throughout the
spacetime outside $r=2M$. We will always
define the shell's trajectory by the value of its $r$-coordinate as
$r=R_s$, however we will treat $R_s$ as a function of either $u$ or
$v$ depending on convenience. The matching conditions are
\begin{align}
&dV - dU = 2dR_s ~~;~~ dv - du = \frac{2dR_s}{(1-2M/R_s)}\,,
  \nonumber\\  
&dUdV = \left(1-\frac{2M}{R_s}\right)dudv \,,
\la{2eq4}
\end{align}
with the differentials understood to be along the trajectory. For a
timelike trajectory, these can be used to write the equation of the
trajectory in the following alternative forms which will be useful
later 
\be
2R_s^\prime(1-U^\prime) = U^{\prime2} - \left(1-\frac{2M}{R_s}\right)\,,
\la{2eq5}
\ee
\be
2\dot R_s(1-\dot V) = \left(1-\frac{2M}{R_s}\right) - \dot V^2\,,
\la{2eq6}
\ee
where the prime and dot denote derivatives with respect to $u$ and $v$
respectively, along the trajectory.
Once the trajectory is specified, these equations completely determine
the metric both inside and outside and also allow us to relate the
coordinates in the two regions.

We next consider the quantization of the scalar field in this
spherical geometry. Rigorously speaking, one needs to introduce the
spherical harmonics $Y_{lm} (\theta,\phi)$ and separate out the
angular dependence of the modes. The resulting expressions will lead
to  a VEV of the stress tensor which cannot be evaluated
analytically. To make progress, we shall concentrate only on the
s-wave component of the scalar field and reduce the  problem to one in
the $r-t$ plane. In such a 2-dimensional context, one can use the
techniques of conformal field theory to evaluate the VEV of
stress-tensor. We further ignore the effects of the Schwarzschild
potential barrier which arises in the exterior even for the s-wave
component. (See \Cites{brout,page,sanchez} for a detailed discussion
of the accuracy of these approximations, and \Cite{visser-bh-ess} for
elementary arguments supporting the s-wave approximation.) The scalar
field is therefore taken to satisfy the 2-dimensional Klein-Gordon
equation both inside and outside the shell, i.e.,  
\begin{align}
\p_U\p_V \vphi &= 0 \,, ~~ {\rm (interior)}\,,\nonumber\\
\p_u\p_v \vphi &= 0 \,, ~~ {\rm (exterior)}\,.
\la{KGeqn}
\end{align}
We shall now briefly recall the procedure for quantizing the scalar
field in the 1+1 spacetime and collect together the relevant formulas.
The general solution of the Klein-Gordon equation in the exterior
region is  
\be
\vphi(u,v) = f(v) + \xi(u)\,,
\la{2eq7}
\ee
which after matching at the shell becomes
\be
\vphi(U,V) = f(v(V)) + \xi(u(U)) \equiv \ti{f}(V) + \ti{\xi}(U)\,,
\la{2eq8}
\ee
in the interior region. Imposing the reflection condition that the
solution vanish at the center $r=(V-U)/2=0$ then gives
\be
\vphi = \ti{f}(V) - \ti{f}(U) = f(v(V)) - f(v(U))\,,
\la{2eq9}
\ee
in the interior, where $v(U)$ is the function $v(V)$ evaluated at
$V=U$. We think of these solutions as  wave-packets and
study each frequency mode separately. Denote by $\vphi^{\rm in}_\om$
modes which are positive frequency with respect to Schwarzschild time
on past null infinity $\Cal{I}^-$. These modes will then define a
vacuum (see below) which corresponds to the Minkowski vacuum on 
$\Cal{I}^-$. The relevant in-falling part of $\vphi^{\rm in}_\om$ is 
\be
\vphi^{\rm in}_\om \sim \frac{1}{\sqrt{4\pi\om}}e^{-i\om v}\,, ~~({\rm
  on\,\,} \Cal{I}^-)\,,
\la{2eq10}
\ee
and is normalized with respect to the Klein-Gordon norm
\be
\left(f_1,f_2\right)_{(v)} = \int_{-\infty}^{\infty}{dv
  f_2^\ast(i\overleftrightarrow{\p_v})f_1}\,,
\la{2eq11}
\ee
and similarly with $u$. The previous discussion on matching and
reflection conditions shows that the outgoing $u$-dependent part of
these modes, which reaches future null infinity $\Cal{I}^+$, is given
by 
\be
\vphi^{\rm in}_\om \sim - \frac{1}{\sqrt{4\pi\om}}e^{-i\om G(u)}\,,
~~({\rm on\,\,} \Cal{I}^+)\,,
\la{2eq12}
\ee
where the function $G(u)$ is defined by the following chain :
An ingoing mode labelled by $v=v_{\rm ent}$ behaves like $\sim
e^{-i\om v_{\rm ent}}$ in the exterior. It enters the shell at the
event $\Cal{P}_{\rm ent}$ and now behaves like $\sim e^{-i\om v_{\rm 
    ent}(V)}$ in the interior. The function $v_{\rm ent}(V)$ is
determined by the matching conditions at $\Cal{P}_{\rm ent}$. At
reflection the mode becomes $\sim e^{-i\om v_{\rm ent}(U)}$, and after
exiting the shell at the event $\Cal{P}_{\rm exit}$ it becomes $\sim 
e^{-i\om v_{\rm ent}(U(u))}$. The function $G(u)$ is hence given by
\be
G(u) = v_{\rm ent}(U(u))\,.
\la{2eq13}
\ee
The scalar field is quantized by defining a vacuum state \state{in}
with respect to these modes by writing the field operator as
\be
\vphi = \int_0^\infty{d\om\left(a_\om \vphi^{\rm in}_\om + {\rm
    h.c.\,} \right)}\,,   
\la{2eq14}
\ee
where ``h.c.'' stands for Hermitian conjugate, and the $a_\om$ are
annihilation operators for the state \state{in}, 
\be
a_\om\state{in} = 0\,.
\la{2eq15}
\ee
We will work throughout in the Heisenberg picture, so that the state
does not evolve, and hence all expectation values must be computed in
the state \state{in}. 

A key point is that on $\Cal{I}^+$, the field operator has
modes which are positive frequency with respect to $G(u)$, not $u$. So
the state \state{in} is the vacuum of the $G$-modes, or the
``$G$-vacuum''. Since $G(u)$ is a nonlinear function of $u$  in
general, this $G$-vacuum will contain particles corresponding to the
``$u$-vacuum'', defined as the state annihilated by modes which are
positive frequency with respect to $u$. 
We will denote this vacuum by the state \state{out}, annihilated by
operators $b_\lam$ corresponding to modes $\vphi^{\rm out}_\lam$, such
that on $\Cal{I}^+$, the relevant outgoing part of $\vphi^{\rm
  out}_\lam$ is  
\be
\vphi^{\rm out}_\lam \sim \frac{1}{\sqrt{4\pi\lam}}e^{-i\lam u}\,,
~~({\rm on\,\,}  \Cal{I}^+)\,,
\la{2eq16}
\ee
and it should be clear that the corresponding in-falling part on
$\Cal{I}^-$ is
\be
\vphi^{\rm out}_\lam \sim -\frac{1}{\sqrt{4\pi\lam}}e^{-i\lam h(v)}\,,
~~{\rm on\,\,}  \Cal{I}^-\,,
\la{2eq17}
\ee
where if $u=h(v)$ then $v=G(u)$. The field operator \vphi\ can be
expanded in the $\vphi^{\rm out}_\lam$ modes as 
\be
\vphi = \int_0^\infty{d\lam\left(b_\lam \vphi^{\rm out}_\lam + {\rm
    h.c.\,} \right)}\,,    
\la{2eq18}
\ee
and a Bogolubov transformation relates the $a_\om$ with the
$b_\lam$. (For definitions and properties of Bogolubov transformations
see \Cite{bir-dav}. We will not require these details.) To summarize,
we have 
\be
\vphi = \int_0^\infty{d\om\left(a_\om \vphi^{\rm in}_\om + {\rm
    h.c.\,} \right)} = \int_0^\infty{d\lam\left(b_\lam \vphi^{\rm
    out}_\lam + {\rm  h.c.\,} \right)}\,,  
\la{2eq19}
\ee
where
\begin{align}
&\vphi^{\rm in}_\om\sim\frac{1}{\sqrt{4\pi\om}}\left\{ 
\begin{array}{l}
e^{-i\om v}~~~~~~~~~ ({\rm on\,\,} \Cal{I}^-) \\
-e^{-i\om G(u)}~~~ ({\rm on\,\,} \Cal{I}^+)
\end{array}\right . \nonumber\\
&\nonumber\\
&\vphi^{\rm out}_\lam\sim\frac{1}{\sqrt{4\pi\lam}}\left\{
\begin{array}{l}
e^{-i\lam u}~~~\,~~~~~ ({\rm on\,\,} \Cal{I}^+) \\
-e^{-i\lam h(v)}~~~ ({\rm on\,\,} \Cal{I}^-)\,,
\end{array}\right .
\la{2eq20}
\end{align}
where, if $v=G(u)$ then $u=h(v)$, and we have $G(u)=v_{\rm ent}(U(u))$
as discussed earlier.

It is now possible to write down an expression for the time dependent
part of the VEV of the 
stress tensor at any stage during the collapse. This is given by (see
Appendix A for details of the notation and derivation)
\begin{align}
\avg{T_{uu}}^{\rm traj}_G(u) \equiv \frac{1}{12\pi}
\left(\frac{dG}{du}\right)^{1/2} \p^2_u
\left(\frac{dG}{du}\right)^{-1/2} = \avg{T_{uu}}^{\rm ren}_G -
\avg{T_{vv}}^{\rm ren}_G\,, 
\la{2eq43}
\end{align}
which is implicitly trajectory dependent. It can be easily verified that 
in the limit when the surface of the
collapsing object approaches its Schwarzschild radius, 
the function $G(u)$ has the asymptotic form 
 $G(u)\propto-4Me^{-u/4M}$
independent of the details of the trajectory.
Hence 
\be
\left(\frac{dG}{du}\right)^{1/2} \p^2_u
\left(\frac{dG}{du}\right)^{-1/2} = \frac{1}{64M^2}\,,
\la{2eq38}
\ee
thereby leading to a flux 
which asymptotes to $(\pi/12)T_H^2$ at late stages of
the collapse. For later use, we
write $\avg{T_{uu}}^{\rm traj}_G(u)$ in an alternative form. First
note that since $G(u)=v_{\rm ent}(U(u))$, differentiating using the 
chain rule gives $G^\prime=dG/du = (dv_{\rm ent}/dV)(dU/du) =
U^\prime/\dot V$ with the understanding that all functions of $v$ such
as $\dot V=dV/dv$ and its derivatives (see \eqn{2eq44} below) are to
be evaluated at $v=v_{\rm ent}(U(u))$. Using the identity 
\be
F^{1/2}\p^2_xF^{-1/2} = -\frac{1}{4}\left[\,2\p^2_x\ln F - (\p_x\ln F)^2\,
\right] \,,
\la{2eq35}
\ee
that is valid for any function $F(x)$, where $\p^2_x\ln F = \p_x[(\p_xF)/F]$, and
repeatedly using the chain rule, it is not hard to show that,   
\be
\avg{T_{uu}}^{\rm traj}_G(u) =
\frac{1}{12\pi} \left((U^\prime)^{1/2}  
\p^2_u (U^\prime)^{-1/2} - \left. \frac{U^{\prime2}}{\dot V^2}(\dot
V)^{1/2} \p^2_v (\dot V)^{-1/2}\right|_{v=v_{\rm ent}(U(u))} \right) \,,
\la{2eq44}
\ee
Due to the presence of the chain
 of functional dependences
 $v_{\rm ent}(U(u))$, explicit
calculations of the VEV will require knowing the functions $v(V)$ and
$U(u)$. In principle, once a trajectory $r=R_s(u)$ or $r=R_s(v)$ is
specified, the matching conditions \eqns{2eq4} can be solved to get
$U(u)$ and $V(v)$. In practice, it is difficult to get both $U(u)$ and
$V(v)$ in closed form, and the matter is further complicated by the
inversion required to get $v_{\rm ent}(V)$ at the point of entry
$\Cal{P}_{\rm ent}$. The analysis can be done for a few
special cases, and Brout et al. \cite{brout,brout-appD} for example
have an explicit calculation (in parametric form) of the
time-dependent flux outside a collapsing star whose internal geometry
is governed by a homogeneous dust (\Cite{brout}, Appendix D). The
asymptotic behaviour of attaining a constant flux is also displayed
very nicely in that calculation (\Cite{brout}, Figure D.1).
\subsection{Trajectory with null in-falling phase without horizon
  formation} 
\label{null}
\noindent
Regardless of calculational difficulties, the general expression
\eqref{2eq43} highlights an important point : The presence of a flux
of particles at large distances is governed entirely by the
local dynamics (in outgoing time $u$) of the collapsing object. A flux
will arise even in the situation where the object starts at some
radius $R_0$ and collapses to a final smaller radius $R_f>2M$ 
\emph{without forming a black hole}, although the spectrum will not be
thermal. Physically one would of course expect that the flux will only
arise during the collapsing phase of the trajectory, and we will
explore this feature next.  We will first
consider a situation in which a shell collapses from $R_0$ to $R_f$
along a \emph{null} trajectory $v=\,$const., after which it remains
static at $r=R_f$. Although this is a rather unrealistic situation, it
does admit an exact calculation of the VEV of the stress tensor  at
any time. We will therefore present the calculations for this case
first. Following this, in the next section, we will give results for a
toy trajectory which remains timelike throughout and also does not
form a horizon. In this case we will be able to provide order of
magnitude estimates for the second term in \eqn{2eq44}, while the
first term will be exactly calculable. 

Recall that we are working in a geometry described by
\eqns{2eq1}-\eqref{2eq6}. Our example trajectory $r=R_s$ comprises
three phases -- 
\begin{eqnarray}
u<0 &:& R_s(u) = R_0 = \text{const.}\nonumber\\
0<u<u_1 &:& v=\,\text{const.},\  V=\, \text{const.}\nonumber\\
u>u_1 &:& R_s(u)=2M/(1-\eps^2)=\,\text{const.}\nonumber
\end{eqnarray}
with $u_1$ defined in terms of \eps\ as described below.
We will ignore the derivative
discontinuities at $u=0$ and $u=u_1$, since these can be smoothed away
if needed, and in any case will not appear when we consider the more
realistic timelike trajectory in \Sec{timelike} below. The
parametrization of the final radius is chosen to ease comparison with
this timelike case. For convenience we define the ``tortoise''
function $x(r)$ as 
\be
x(r) \equiv r + 2M\ln\left(r/2M-1\right) ~~;~~ r>2M\,.
\la{toy-eq4}
\ee
Using the trajectory equations \eqref{2eq4}, fixing some constants of
integration and ensuring continuity at the transition events $u=0$ and
$u=u_1$, the functional form of the trajectory can be shown to be as
follows: For $u<0$ (Phase (1)), we get
\begin{equation}
 U(u) = \left(1-\frac{2M}{R_0}\right)^{1/2}u \,-\,
2R_0 ~~;~~ V(v) = \left(1-\frac{2M}{R_0}\right)^{1/2}v \,-\,
2x(R_0)\left(1-\frac{2M}{R_0}\right)^{1/2} \,, 
\la{toy-eq7} 
\end{equation}
For $0<u<u_1$ (Phase (2)), we get
\begin{equation}
 v=2x(R_0)~~;~~ V=0~~;~~ U = -2R_s ~~;~~ u =
-2x(R_s) + 2x(R_0)\,, 
\la{toy-eq8} 
\end{equation}
For $u>u_1$ (Phase (3)) we get
\begin{equation}
 U(u)=\eps(u-u_1) - 4M/(1-\eps^2) ~~;~~ V(v) 
=  \eps(v-2x(R_0))  ~~;~~ u_1 = 2x(R_0) -
2x(2M/(1-\eps^2))\,. 
\la{toy-eq12}
\end{equation}
The Penrose diagram for this trajectory is shown in \fig{pen-toytraj}.
\begin{figure}[t]
\centering
\includegraphics[height=0.35\textheight]{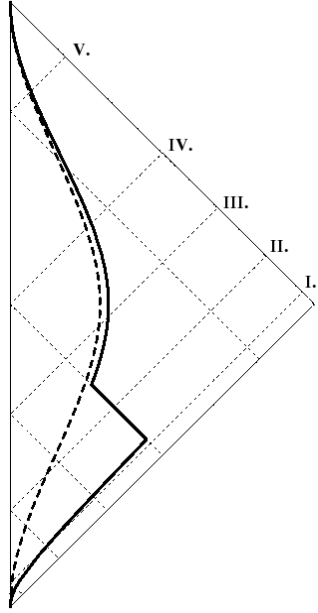}
\caption{Penrose diagram for the trajectory described by
  \eqns{toy-eq7}-\eqref{toy-eq12}. We have set $R_0=12M$ and
  $\eps^2=0.1$. The thick solid line is the shell trajectory, and the
  dashed line is the timelike surface $r=2M$. The interior of the
  shell is Minkowski spacetime while the exterior is
  Schwarzschild. The five types of null rays relevant for the stress
  tensor VEV calculation are also shown.}    
\la{pen-toytraj}
\end{figure}
There are five types of null rays (which we label I,II, ...V) relevant
for the calculation of $G(u)$ and hence the stress tensor VEV,
characterized by the locations of entry and exit events $\Cal{P}_{\rm
  ent}$ and $\Cal{P}_{\rm ex}$. These correspond to:  {(I)}
$\Cal{P}_{\rm ent}$ and $\Cal{P}_{\rm ex}$ in phase (1). 
 {(II)} $\Cal{P}_{\rm ent}$ in phase (1), $\Cal{P}_{\rm ex}$
  in phase (2).
 {(III)} $\Cal{P}_{\rm ent}$ in phase (1), $\Cal{P}_{\rm ex}$
  in phase (3).
 {(IV)} The single ingoing ray $v=2x(R_0)$ or $V=0$, which
  skims the trajectory and finally enters the shell at $u=u_1$, with
  $\Cal{P}_{\rm ex}$ in phase (3).
 {(V)} $\Cal{P}_{\rm ent}$ and $\Cal{P}_{\rm ex}$ in phase (3). 
For type (I) rays, we have
\be
v_{\rm ent}(V) = \left(1-\frac{2M}{R_0}\right)^{-1/2}V + {\rm const.}
~~;~~ U(u) =  \left(1-\frac{2M}{R_0}\right)^{1/2}u + {\rm const.}\,,
\la{toy-eq13}
\ee
and hence $G(u)=u+\,$const. which leads to a vanishing flux. 
For type (II) rays, we have
\be
v_{\rm ent}(V) = \left(1-\frac{2M}{R_0}\right)^{-1/2}V + 2x(R_0) ~~;~~
e^{-u/4M} = e^{(R_s-R_0)/2M}\frac{(R_s-2M)}{(R_0-2M)} ~~;~~ R_s = -U/2 \,,
\la{toy-eq14}
\ee
and hence 
\be
G(u) = \left(1-\frac{2M}{R_0}\right)^{-1/2}U(u) + 2x(R_0)\,,
\la{toy-eq15}
\ee
with $U(u)$ given implicitly by the last two equations in
\eqref{toy-eq14}. We will soon return to the flux obtained from this
form of $G(u)$. Note that type (II) rays all occur in $0<u<u_1$.
For type (III) rays, we have
\be
v_{\rm ent}(V) = \left(1-\frac{2M}{R_0}\right)^{-1/2}V + {\rm const.}
~~;~~ U(u) =  \eps u + {\rm const.}\,,
\la{toy-eq16}
\ee
which gives $G(u) = \eps(1-2M/R_0)^{-1/2}u+\,$const. which is also
linear in $u$ and hence gives a vanishing flux.
Finally, type (IV) and (V) rays have
\be
v_{\rm ent}(V) = \eps^{-1}V + {\rm const.}
~~;~~ U(u) =  \eps u + {\rm const.}\,,
\la{toy-eq17}
\ee
also leading to $G(u)=u+\,$const. and hence a vanishing flux. 

The only nonzero flux therefore arises for type (II) rays, in the
interval $0<u<u_1$. For $\avg{T_{uu}}^{\rm traj}(u)$ defined in
\eqn{2eq43}, a straightforward calculation leads to
\be
\avg{T_{uu}}^{\rm traj}(u) = \frac{1}{48\pi}\frac{M}{R_s(u)^3}\left( 2 -
\frac{3M}{R_s(u)} \right)\,, ~~ 0<u<u_1\,,
\la{toy-eq21}
\ee
and zero otherwise, where $R_s(u)$ in the given range is implicitly
determined through
\be
e^{-u/4M} = e^{(R_s-R_0)/2M}\frac{(R_s-2M)}{(R_0-2M)}\,.
\la{toy-eq22}
\ee
The behaviour of $\avg{T_{uu}}^{\rm traj}(u)$ normalized by the
Hawking value $(\pi/12)T_H^2$, is shown
in \fig{back-toy}, for $R_0=12M$ and two values of $\eps$. The
step-like rise and fall are due to the derivative discontinuities
mentioned  earlier, and do not occur in the timelike case which we
will study next. The important feature to note is that for small
\emph{but nonzero} $\eps$, the flux attains the asymptotic Hawking
value but eventually falls to zero. \eqns{toy-eq21} and
\eqref{toy-eq22} show that the time taken to attain the
Hawking value is governed by $R_0$, which is physically clear since
the exponential redshift of the outgoing modes occurs only near
$r=2M$, so that for larger $R_0$, the shell spends more time radiating
a smaller non-thermal flux. This argument can be easily verified
numerically as well. 
\begin{figure}[t]
\centering
\includegraphics[height=0.2\textheight]{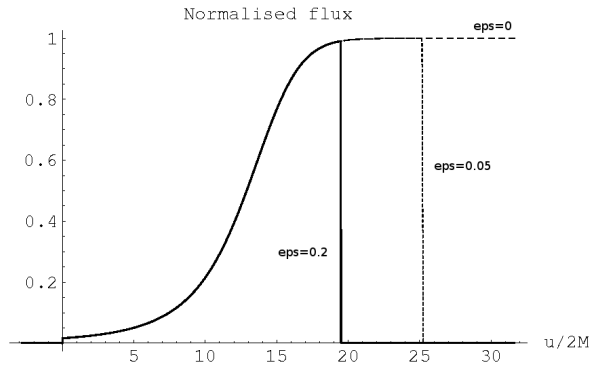}
\caption{The behaviour of $\avg{T_{uu}}^{\rm traj}(u)$ normalized by
  the Hawking value $(\pi/12)T_H^2$, for the trajectory studied in
  \Sec{null}. We have set $R_0=12M$. The solid and dotted lines
  show the flux for two values of $\eps$ as labelled. The dashed
  line is the asymptotic Hawking flux corresponding to $\eps=0$.}    
\la{back-toy}
\end{figure}
%
\subsection{Timelike trajectory without  horizon formation}
\label{timelike}
\noindent
We will now turn to a somewhat more realistic trajectory which is
continuous, differentiable and timelike at all times, and which also
does not form a horizon. The specific example we choose is a
trajectory which remains fixed at $r=R_0$ until $u=0$, and
\emph{asymptotically} (as $u\to\infty$) approaches a final radius
$R_f=2M/(1-\eps^2)>2M$. It turns out that this can be easily
accomplished by appropriately choosing a function $U^\prime =
dU/du$. [Since this calculation is for illustrative purposes, we will
  not worry about the shell dynamics needed to obtain the behaviour
  described below.] Consider then the following prescription for
$U^\prime$, 
\be
U^{\prime} = \eps + \bigg(\left(1-\frac{2M}{R_0}\right)^{1/2} - \eps
\bigg) e^{-\alpha(u)} \equiv \eps + Ae^{-\alpha(u)}\,,  
\la{2eq45}
\ee
where $0<\eps<(1-2M/R_0)^{1/2}$ is a fixed constant,
\be
\alpha(u)=\frac{\theta(u)}{4M}\int_0^u{d\ti{u} h(\ti{u})}\,,
\la{2eq46}
\ee
where $\theta(u)$ is the Heaviside step function and $h(u)$ is chosen
to have the following asymptotic behaviour
\begin{align}
h(u) = 0\,, u\leq0~~&;~~ h^\prime(u)=0\,, u\leq0\,, \nonumber\\
h(u\to\infty)\to1~~&;~~ h^\prime(u\to\infty)\to0\,,
\la{2eq47}
\end{align}
and we require the asymptotic values for $h$ and $h^\prime$ to be
achieved exponentially fast, with a time scale determined by $M$. An
example of a function $h(u)$ which meets these requirements is
$h(u)=\theta(u)\tanh(\kappa^2u^2)$ for some constant $\kappa$. The
condition that the trajectory remain timelike throughout reduces to
$-2R_s^\prime<U^\prime$ or $(1-2M/R_s)<U^\prime$, which will in
general impose a restriction on the allowed values of the timescale
$\kappa$ for a given starting radius $R_0$. Physically, if the shell
starts from rest at a larger radius $R_0$ then it will take a longer
time (at subluminal velocities throughout) to reach radii where the
asymptotic exponential approach to the final radius begins. Hence a
larger $R_0$ will imply a \emph{smaller} $\kappa$. We have checked
numerically that for $h(u)=\theta(u)\tanh(\kappa^2u^2)$ with $R_0$
significantly larger than $2M$ (say $R_0\gtrsim4M$), the largest
allowed value of  $\kappa$ is $\kappa_{\rm max}\sim M^{-1}(M/R_0)^b$
with $b\approx2$ (and a very weak dependence on $\eps$),
and in general we can expect $b>0$. The Penrose diagram for such an
``asymptotic'' trajectory is shown in \fig{pen-asymp}.
\begin{figure}[t]
\centering
\includegraphics[height=0.35\textheight]{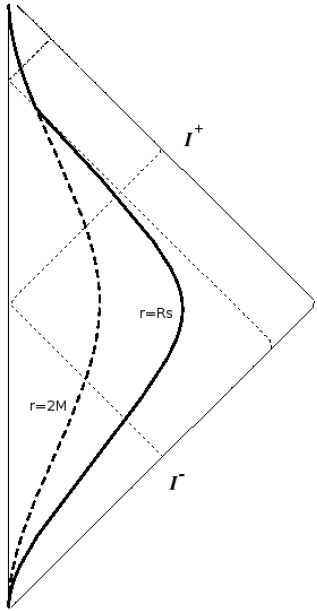}
\caption{Penrose diagram for the ``asymptotic'' trajectory described
  by \eqns{2eq45}-\eqref{2eq47}. We have set $R_0=5M$, $\eps=10^{-2}$
  and $2M\kappa=0.075$. The thick solid line is the shell trajectory,
  and the dashed line is the timelike surface $r=2M$. The interior of
  the shell is Minkowski spacetime while the exterior is
  Schwarzschild. The two null rays mark (by their entry points)
  the beginning of the infall phase and the approximate beginning of
  the asymptotic phase.}     
\la{pen-asymp}
\end{figure}

The reason for this somewhat convoluted prescription is
that we want to simplify the VEV calculations, which involve
derivatives with respect to $u$, and it is therefore convenient to
parametrize the trajectory using $u$. To see that the required
behaviour for the trajectory $R_s(u)$ is reproduced by this
$U^\prime$, one analyses \eqn{2eq5} which is reproduced below 
\be
2R_s^\prime(1-U^\prime) = U^{\prime2} - \left(1-\frac{2M}{R_s}\right)\,.
\la{2eq48}
\ee
The behaviour for $u\leq0$ is obtained correctly. For $u>0$, 
consider the asymptotic regime (large $u$) where $h\approx1$ and
$h^\prime\approx0$. In this regime $U^{\prime2}$ approaches $\eps^2$
from above, forcing $(1-2M/R_s)$ to also approach $\eps^2$
asymptotically. In Appendix \ref{appendix1} we show
that the asymptotic behaviour for the trajectory is 
\be
1-\frac{2M}{R_s} = \eps^2 + f(u)~~;~~ e^{-u/4M}\ll\eps^2 e^{-1/\kappa
  M}\,, 
\la{2eq49}
\ee 
where $f(u)$ is an exponentially decaying function, the exact form of
which depends on the value of \eps. Note that $\eps$ is fixed and we
are not taking a $\eps\to0$ limit. We then see that the required
behaviour is being reproduced. 

\begin{figure}[t]
\centering
\subfigure[Varying
  $\eps$]{\includegraphics[width=.45\textwidth]{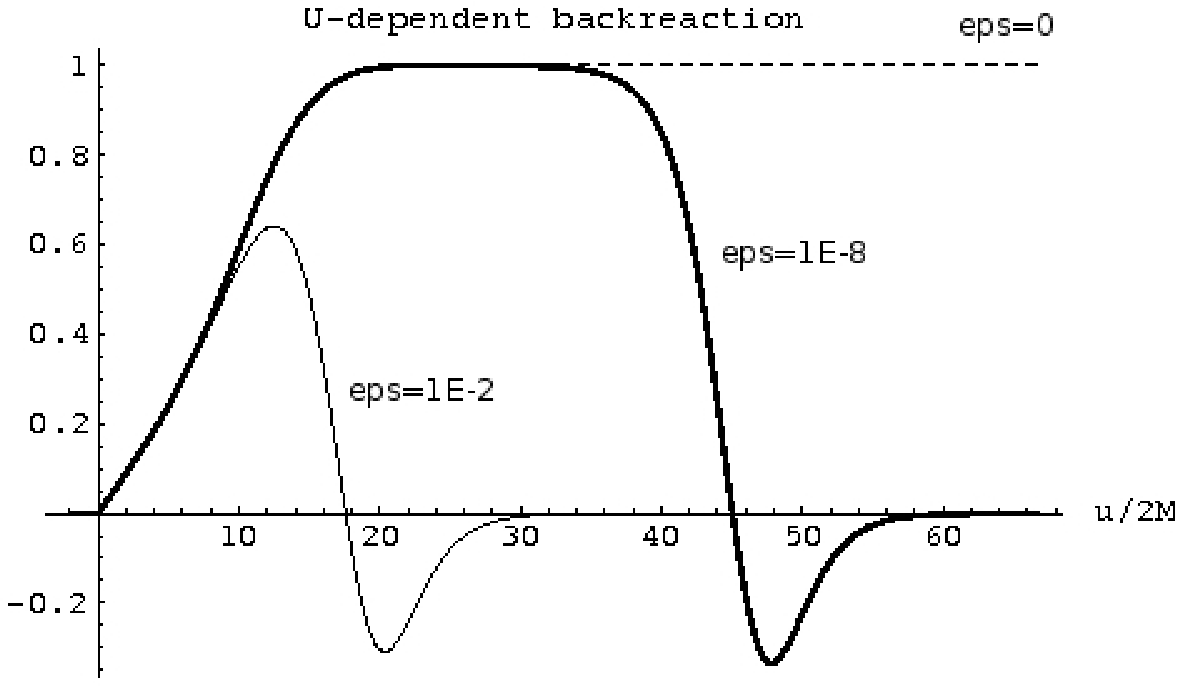}   
\label{backUa}}    
\hspace{.05\textwidth} 
\subfigure[Varying
  $\kappa$]{\includegraphics[width=.45\textwidth]{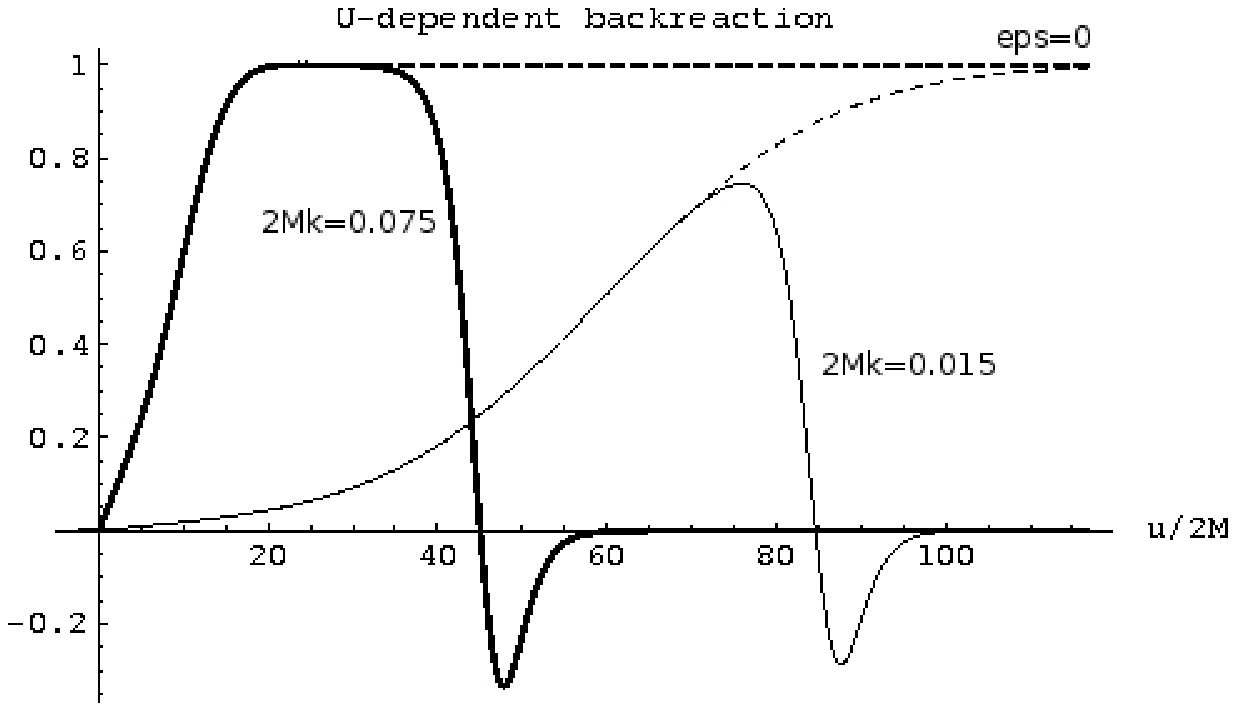}   
\label{backUb}}   
\caption{Behaviour of the $U$-dependent backreaction term
  $(8M)^2[(U^\prime)^{1/2}\p^2_u (U^\prime)^{-1/2}]$, for
  $h(u)=\theta(u)\tanh(\kappa^2u^2)$, with $R_0=5M$. Panel (a) :
  $2M\kappa=0.075$ is fixed. The thin solid curve has $\eps=10^{-2}$,
  the thick solid curve has  $\eps=10^{-8}$, while the dashed curve is
  the standard case with $\eps=0$. Panel (b) : For the solid curves
  $\eps=10^{-8}$ is fixed, while the dashed curves have $\eps=0$. The
  thick curves are for $2M\kappa=0.075$ and the thin curves for
  $2M\kappa=0.015$. See text for discussion.} 
\label{backU}
\end{figure}

As before, we wish to evaluate $\avg{T_{uu}}^{\rm traj}_G(u)$  at
any given time along this trajectory. The first term in \eqn{2eq44} 
involving derivatives of $U^\prime$, can be easily computed exactly
using \eqn{2eq45} and gives
\be
\frac{1}{12\pi} (U^\prime)^{1/2}\p^2_u (U^\prime)^{-1/2} =
\frac{1}{48\pi} \frac{1}{1+(\eps/A)e^{\alpha(u)}} \left[
  \alpha^{\prime2} \left(
  \frac{1-2(\eps/A)e^{\alpha(u)}}{1+(\eps/A)e^{\alpha(u)}} \right)
  +   2\alpha^{\prime\prime}\right]  \,,
\la{2eq50}
\ee 
It is interesting to note that in the case $\eps=0$, i.e. for
a standard trajectory which approaches $r=2M$ as 
$u\to\infty$, this term is positive definite and gives rise to the
Hawking flux $(\pi/12)T^2_H$, which can be easily checked by setting
$\eps=0$ in \eqn{2eq50}. The behaviour of
$(8M)^2[(U^\prime)^{1/2}\p^2_u (U^\prime)^{-1/2}]$ for
$h(u)=\theta(u)\tanh(\kappa^2u^2)$ with $R_0=5M$, is shown in  
\fig{backU}. Panel (a) shows curves for a fixed value of $\kappa$ with 
$2M\kappa=0.075$ (which ensures a timelike trajectory). The two solid 
lines correspond to $\eps=10^{-2}$ and $\eps=10^{-8}$, while the
dashed line corresponds to the standard case with $\eps=0$, with the
asymptotic value corresponding to the Hawking flux. In panel (b), the
solid lines correspond to a fixed value of $\eps=10^{-8}$, with
$2M\kappa=0.075$ and $2M\kappa=0.015$. The $\eps=0$ curves for these
values of $\kappa$ are also shown as dashed lines.

We see that the initial behaviour of the $U$-dependent term is
practically identical to the standard $\eps=0$ case. Eventually the
term involving $\alpha^\prime$ in \eqn{2eq50} becomes negative,
leading to negative values for the right hand side at least for the chosen
$h(u)$. While this sign change might not be a generic feature, the
\emph{late time} behaviour ($u\gg\kappa^{-1}\ln(1/\eps^2)$, see
\eqn{2eq49}) is generically seen to be an exponential decay $\sim
(e^{1/\kappa M}/\eps M^2)e^{-u/4M}$, which directly follows from
\eqn{2eq50}. In Appendix \ref{appendix6} we discuss some additional
features of the behaviour of this $U$-dependent term, which allow us
to place a bound on the mass loss due to backreaction for this type of
trajectory. 

We now turn to the second term in \eqn{2eq44}, which involves
derivatives of $\dot V$, and an evaluation at $v=v_{\rm
  ent}(U(u))$. Unfortunately, unlike \eqn{2eq50} for the first
term, it is not possible here to give an analytic expression valid at
arbitrary times. The problem arises mainly because of the function
$v_{\rm ent}(V)$ which --- in general --- has a complicated
form. Recall that this function is determined by the matching
conditions on the trajectory at the point of entry $\Cal{P}_{\rm
  ent}$, and the evaluation is at $V=U(u)$ where $U(u)$ is determined
by the exit point $\Cal{P}_{\rm exit}$. Despite the lack of exact
expressions, it turns out that we can make order of magnitude
statements by realizing that we are essentially dealing with two kinds
of ingoing rays characterized by the locations of $\Cal{P}_{\rm ent}$
and $\Cal{P}_{\rm exit}$ on the trajectory :
\begin{enumerate}
\item $\Cal{P}_{\rm ent}$ occurs for $u_{\rm ent}>0$ but before the
  asymptotic phase so that $u_{\rm
    ent}\lesssim\kappa^{-1}\ln(1/\eps^2)$.  
\item $\Cal{P}_{\rm ent}$ (and hence $\Cal{P}_{\rm exit}$) occurs in
  the asymptotic phase.
\end{enumerate}
We ignore rays which enter at $u_{\rm ent}<0$, since it is easy to
show that $v_{\rm ent}(V)\propto V$ in this case, giving $\dot
V=\,$const, and hence the second term will be exactly zero. For the
choice of parameters used in \fig{backUa} (with $\eps=10^{-8}$), and
fixing one constant so that $V_{\rm traj}(u=0)=0$,
numerically one finds that the ray which enters at $u_{\rm ent}=0$
will exit at $u_{\rm ex}\approx14M$, and hence
the second term in \eqn{2eq44} will vanish for $u<14M$. In general if
$V_{\rm traj}(u=0)=0$ then $u_{\rm ex}(u_{\rm ent}=0)$ satisfies
\be
\eps u_{\rm ex} + A\int_0^{u_{\rm ex}}{e^{-\alpha(u)}du}=2R_0\,.
\la{2eq50-extra1}
\ee
For rays of type (1), if $u_{\rm ent}\lesssim\kappa^{-1}$, the
requirement that the trajectory be well-behaved in the proper frame of 
the shell, allows us to argue that the second term in \eqn{2eq44} must
be small. For $\kappa^{-1}\lesssim u_{\rm
  ent}\lesssim\kappa^{-1}\ln(1/\eps^2)$, the same arguments show this
term to be at most $\Cal{O}(M^{-2})$ and  changing on a
timescale $\Cal{O}(M)$  (see Appendix \ref{appendix3} for
details). For type (2) rays we need to work entirely in the
asymptotic regime, in which the function $v_{\rm ent}(V)$ turns out to
be approximately linear, and as we show in Appendix \ref{appendix2}
the contribution from the second term of \eqn{2eq44} also decays
exponentially. The details depend on the value of \eps, but the
typical behaviour is 
\be
\left. \frac{U^{\prime2}}{\dot V^2}(\dot
V)^{1/2} \p^2_v (\dot V)^{-1/2}\right|_{v=v_{\rm ent}(U(u))} \sim
\frac{1}{\eps^2M^2}e^{1/\kappa M} e^{-\Cal{K}u/4M} ~~;~~
\Cal{K}=\Cal{O}(1) ~~;~~ e^{-u/4M}\ll\eps^2e^{-1/\kappa M}\,,
\la{2eq51}
\ee 
where $u=u_{\rm ex}$, the exit time. When $\eps$ is significantly
smaller than unity, this term will dominate over the first term (which
falls like $\sim (e^{1/\kappa M}/\eps M^2)e^{-u/4M}$) \cite{footnote1}.
This analysis demonstrates a result which one should intuitively
expect on physical grounds : When the object is initially static,
there is no outgoing flux. When the object starts to collapse, there
is a flux of at most $\Cal{O}(M^{-2})$ which changes
on a timescale $\Cal{O}(M)$, and when the object
approaches its asymptotically static configuration, this flux
exponentially decays to zero.
\section{The Semiclassical Backreaction}
\la{bkrxn}
\noindent
The presence of an outgoing flux of particles during the gravitational
collapse of a body, immediately tells us that the assumption that the
mass of this body is a constant, cannot be correct. The formal way
to see this is to consider the energy conservation equation
$u^bT{}^a_{b;a}=0$ at large distances from the collapsing object,
assuming the Schwarzschild exterior (which is asymptotically
flat). Taking the 4-velocity $u^b$ to be that of an observer at rest
at fixed $(r,\theta,\phi)$, and integrating over a three volume of
radius $r$, we have 
\be
\frac{dM}{dt} = \oint_{r}{d^2S\,T{}^r_t} = 4\pi r^2
T{}^r_t = {}^{(2d)}T{}^r_t  = - \avg{T_{uu}}^{\rm traj}_G \equiv
-L_H\,. 
\la{3eq1}
\ee
The first equality follows from integrating the conservation equation,
with $d^2S = r^2\sin\theta d\theta d\phi$; the second follows from
spherical symmetry; the third follows from assuming that the angular
components of $T_{ab}$ vanish, so that the 4-dimensional values of
$T_{tt}$, $T_{rt}$ and $T_{rr}$ are rescaled versions of their
2-dimensional counterparts; the fourth equality follows from
replacing ${}^{(2d)}T_{ab}$ by its semiclassical value computed
in Appendix A. The last definition is for later ease of notation.

In principle, this could lead to a situation in which the existence of
an outgoing flux prevents the formation of the event horizon. The
radiative flux decreases the  mass of the collapsing body and thus
decreases the effective radius of the event  horizon which the
collapsing surface is chasing. If the radius of the event horizon
decreases fast enough, the collapsing surface will never be able to
catch up with it and as a result the event horizon will not form. Our
aim now is to analyse this situation more carefully and show that this 
scenario can be ruled out. Backreaction does not prevent the
formation of the event horizon.

In reality, what comes to our rescue is  the following fact:  the mass
loss rate is set by the semiclassical backreaction which is governed
by the ratio $1/M^2$ in \emph{Planckian units}, which makes it a very
small number (e.g., for a solar mass object we have
$M_{\odot}\simeq10^{38}$). It then turns out that under
the assumption that the outgoing flux (and hence mass loss rate) is
small and slowly varying, a straightforward ansatz is enough to
estimate the effect of the backreaction on the background geometry
self-consistently. Brout et al. \cite{brout} present a calculation
(which we sketch in Appendix \ref{appendix4}) which shows
that the following late time scenario is self-consistent (see also
\Cite{massar}, and \Cites{bardeen,york} for the original papers): 
\begin{description}
{\item (a)} The time dependence of the outgoing flux $L_H(u)$ at late
times is determined solely by the \emph{time-dependent} mass $M(u)$,
and is given by $L_H\sim1/M^2$, with $u$ being an
``Eddington-Finkelstein-like'' outgoing coordinate. $L_H$ is assumed
to be small compared to unity (in Planckian units) and slowly
varying. 
{\item (b)} The exterior geometry at large distances, in terms of the
outgoing coordinate $u$, is the outgoing Vaidya solution given by
\be
ds^2_{{\rm ext, large\,} r} = -\left(1-\frac{2M(u)}{r}\right)du^2 -
2dudr + r^2d\Omega^2 ~~;~~ \frac{dM(u)}{du} = -L_H\,.
\la{3eq2}
\ee
{\item (c)} The exterior geometry in terms of an ingoing coordinate
$v$, at any distance, is approximately (i.e. up to terms of order
$\Cal{O}(L_H)$) given by
\be
ds^2_{\rm ext} \approx -\left(1-\frac{2m(v,r)}{r}\right)dv^2 + 2dvdr +
r^2d\Omega^2 \,,
\la{3eq3}
\ee
where the mass function $m(v,r)$ is slowly varying in the entire
exterior, in that  
\be
\frac{\p m}{\p v} = \Cal{O}(L_H) ~~;~~ \frac{\p m}{\p r} =
\Cal{O}(L_H) \,,
\la{3eq4}
\ee
with the transformation between the $u$ and $v$ coordinate being such
that at large distances one recovers the outgoing Vaidya metric
\eqref{3eq2} with $m(v,r)=M(u)$. (See Appendix \ref{appendix4} for
details of the transformation.)
{\item (d)} Self-consistency is demonstrated by showing that the VEV of the
stress tensor  $\avg{T_{uu}}^{\rm traj}(u)$ computed in this
geometry does indeed behave as $\sim M(u)^{-2}(1+\Cal{O}(L_H))$.
\end{description}

The Brout et al. calculation (as acknowledged by those authors) is
only valid in the regime where the flux $L_H$ is small and slowly
varying. There will inevitably be a phase in the collapse when the
mass loss rate becomes significant enough that a perturbative
expansion in $L_H$ is no longer valid. A full fledged calculation of
the backreaction in this regime, to our knowledge, has not yet been
performed.  Our interest however, is only to ask
whether the backreaction can delay the formation of the event horizon to the
extent that it does not form at all. We will not worry about any
significant backreaction effects that occur \emph{after} the horizon
has formed. 

It is important to note that all questions about event horizon
formation \emph{must} be asked in a reference frame where this 
formation occurs in a finite time in the unperturbed collapse. It is
not possible to theoretically settle this issue if one insists on
working entirely in the coordinates used by static observers at large
distances, even though these may be the most natural coordinates to
use, simply because even in the \emph{classical} scenario, event horizon
formation takes an infinite amount of time in these coordinates.  

Consider then a situation in which the unperturbed collapse trajectory
of the shell crosses the radius $r=2M$ (with constant $M$) in finite
proper time with a finite subluminal velocity. The exterior geometry
for the unperturbed collapse is given by the Schwarzschild metric 
\eqref{2eq1}, while the interior is Minkowski spacetime
\eqref{2eq3}. [It is also interesting to consider a
trajectory whose in-falling phase is light-like, an extension as it
were of the trajectory we studied in \Sec{null}. We have analysed
such a trajectory in Appendix \ref{appendix5}.] On parametrizing the 
timelike trajectory using the shell proper time $\tau$ (with
$ds^2|_{\rm traj}=-d\tau^2$), we find that the quantity $dv/d\tau$
remains finite at horizon formation. It is then convenient to
reparametrize the trajectory using the ingoing Eddington-Finkelstein
coordinate $v$, as $(r=\bar R(v),V=\bar V(v))$. Assuming that horizon
formation in this unperturbed case occurs at some time $v=v_0$, we
have 
\be
\bar R(v) - 2M = -k(v-v_0) + \Cal{O}((v-v_0)^2)~,~~ {\rm as\,\,} v\to 
v_0\,, 
\la{3eq5}
\ee
where the constant $k$ can be related to the proper velocity
$\beta_h\equiv-(d\bar R/d\tau)|_{\bar R=2M}$, as $k=2\beta_h^2$ (which  
follows from using the metric \eqref{2eq1} on the trajectory). The
trajectory equation \eqref{2eq6} becomes
\be
2\dot {\bar R}(1-\dot {\bar V}) = \left(1-2M/\bar R(v)\right) -
\dot {\bar V}^2\,,
\la{3eq6}
\ee
(where the dot is a derivative with respect to $v$ along the
trajectory) which shows that $\dot {\bar V}$ also remains finite at 
$\bar R=2M$.  

Let us now incorporate the effects of a backreaction caused by a small
outgoing flux $L_H\ll1$, and ask by what amount is the event horizon
formation delayed. Specifically, let the trajectory now be
$(r=R(v),V=V(v))$ where the interior metric is still Minkowski
spacetime in $(r,V)$ coordinates \eqref{2eq3}. The exterior metric,
following Brout et al., is given by \eqn{3eq3}, and the trajectory
equation \eqref{2eq6} is easily shown to be replaced by 
\be
2\dot R(1-\dot V) = \left(1-2m(v,R(v))/R(v)\right) - \dot V^2\,.
\la{3eq7}
\ee
For the exterior metric \eqref{3eq3}, the event horizon (if it forms)
is the last outgoing null ray $r=r_{\rm eh}(v)$ and satisfies the
outgoing null geodesic equation
\be
\frac{dr_{\rm eh}}{dv} = \frac{1}{2}\left(1 - \frac{2m(v,r_{\rm
    eh}(v))}{r_{\rm eh}(v)} \right) \,.
\la{3eq8}
\ee
Note that this event horizon is distinct from the \emph{apparent
  horizon} $r=r_{\rm ah}(v)$ which is defined as the locus of events
at which $dr/dv=0$ along outgoing null geodesics, so that $r_{\rm
  ah}(v) = 2m(v,r_{\rm ah}(v))$. However as we show in Appendix
\ref{appendix4}, assuming that the flux $L_H$ is small, we get  
\be
2m(v,r_{\rm eh}(v)) = r_{\rm eh}(v)(1+\Cal{O}(L_H))\,.
\la{3eq9}
\ee
so that this distinction is irrelevant. We now use the smallness of
$L_H$ to make the following assumptions, 
which will turn out to be self-consistent at the end of the
calculation. (a) We assume that the \emph{functional form} of the
trajectory is affected only by terms of order $\Cal{O}(L_H)$, so that
\be
R(v) = \bar R(v) \left(1 + \Cal{O}(L_H)\right) ~~;~~ V(v) = \bar V(v)
\left(1 + \Cal{O}(L_H)\right) \,,
\la{3eq10}
\ee
which is reasonable since the trajectory of the event horizon ($r=2M$
in the unperturbed case) is also affected by the same amount (see
\eqns{3eq8}, \eqref{3eq9}); and (b) the event horizon is formed
(i.e. $R(v)=r_{\rm eh}(v)$ is satisfied) at some finite $v=\ti{v}_0$
such that $\ti{v}_0 = v_0 + \delta v$ where $\bar R(v_0)=2M$ and
$\delta v$ is to be determined. We now evaluate the perturbed
trajectory equation \eqref{3eq7} at $v=\ti{v_0}$, linearize around
$v=v_0$ assuming that $\delta v$ is ``small'', and use the unperturbed
equation \eqref{3eq6} to obtain after straightforward algebra,
\be
\delta v = \Cal{O}\left( \frac{2M}{k}L_H\right) = \Cal{O}\left(
\frac{M}{\beta_h^2}L_H\right) \,,
\la{3eq11}
\ee
where $\beta_h$ is the proper velocity with which
the unperturbed trajectory crosses $r=2M$ and $L_H$ is the luminosity
of the thermal radiation. This result is intuitively
clear : The delay in the event horizon formation is governed by the
product of $L_H$ and the timescale of the unperturbed collapse, which
is essentially set by the initial mass of the collapsing
object. 

One might still argue however, that there could in principle exist
\emph{unperturbed} trajectories which slow down sufficiently while
still outside $r=2M$, that $\beta_h$ becomes significantly smaller than
unity. In such a case, it might be possible to delay the event horizon
formation to such an extent that the backreaction itself starts
changing significantly and becomes large, leading to a runaway process
which might exclude the formation of the event horizon. While this
situation cannot be excluded, the following argument does render it
implausible. 
\begin{itemize}
\item First note that there is no semiclassical radiation in static
  geometries. The flux of particles arises when the geometry
  (characterized e.g. by the shell radius) changes by a significant
  amount during the time that an ingoing mode reflects at the center
  and exits the object. For this to occur the trajectory must be
  governed by a timescale \emph{linear} in $M$, since this is the order
  of the time spent by the modes inside the object. For example in the
  ``asymptotic'' trajectory of \Sec{timelike}, the maximum flux is
  achieved for times $u\gg\kappa^{-1}$ when the timescale of the
  function $\alpha(u)$ is set by $M$.
\item Our calculations in \Sec{shellradn}, and also e.g. the
  calculation in Brout et al.'s Appendix D, show that in any stage of
  the trajectory governed by the timescale $M$, the flux is expected
  to be at most $\Cal{O}(1/M^2)$, with the largest value expected only
  as the object approaches $r=2M$. In \Sec{timelike} we also saw that
  in the extreme case when the trajectory is asymptotically slowed
  down to a halt, the flux in fact exponentially decays. One expects
  therefore that if a trajectory is slowed down by some smaller amount
  (i.e. not exponentially), the flux will have to lie between its
  maximum of $\Cal{O}(1/M^2)$, and zero. In particular, slowing down a 
  trajectory cannot \emph{increase} the flux. 
\item In order to completely evaporate the collapsing object before
  the event horizon forms, what we need along the \emph{unperturbed}
  trajectory is to sustain the maximum possible flux for the largest
  possible time. Naively one would want a flux $\Cal{O}(1/M^2)$ (which
  is the maximum possible) sustained for a time $\Cal{O}(M^3)$. The
  previous arguments show that this is unlikely to happen, since there
  is a trade-off between having a significant flux and remaining
  outside $r=2M$ for long enough.   
\end{itemize}
\subsection{Bound on the total radiated mass}
\la{bound}
\noindent
For trajectories such as those discussed in \Sec{shellradn}, and in
fact for any trajectories that have such a form up to any finite time,
we can place concrete bounds on the amount of mass that can be
radiated away. We do this by calculating or estimating the integral
$\Delta M = \int{\avg{T_{uu}}^{\rm traj}du}$. For the trajectory of
\Sec{null}, $\avg{T_{uu}}^{\rm traj}$ is non-zero only in the range
$0<u<u_1$, in which its integral can be explicitly performed to give 
\be
\frac{\Delta M}{M} = \frac{-1}{48\pi}\frac{1}{4M^2}\left[
  \ln(\eps^{-2}) + \ln(1-2M/R_0) +(2M/R_0)-(1-\eps^2) +
  \frac{3}{2}\left\{(2M/R_0)^2 - (1-\eps^2)^2\right\}\right]\,. 
\la{mass-loss-null}
\ee
We expect $R_0/M$ to be significantly larger than unity. Also, since
we are in the semiclassical domain, we can only expect to make
meaningful statements for situations where the final asymptotic
radius is \emph{not closer to} $r=2M$ than one Planck length.
This Planck length bound translates to a lower bound on \eps\ given
by $\eps^2>1/(2M+1)$, so that the maximum mass that can be radiated
away (ignoring numerical factors) is 
\be
\left|\Delta M_{\rm max}\right| \sim M^{-1}\ln(M)\,.
\la{mass-loss-bound-null}
\ee
For the asymptotic trajectory of \Sec{timelike}, the situation is 
trickier since we do not have exact expressions for the full
backreaction, and also because the presence of the timescale
$\kappa^{-1}$ causes a subtle interplay of competing effects. We
show in Appendix \ref{appendix6} however, that even in this case
there is an upper bound on the possible mass loss, which is in fact
\emph{identical} (up to numerical factors) to the expression
\eqref{mass-loss-bound-null}. 
\section{Conclusions}
\la{conclude}
\noindent
Since we have provided a detailed summary of results right at the
beginning of the paper, we shall be brief in this section and will
just review and stress the key conclusions which we have obtained. 

It is possible to explicitly compute the VEV of the stress tensor
around a collapsing shell for any arbitrary trajectory of the
shell. In principle, this calculation should  be done in (1+3)
dimensions but the results based on s-waves can be obtained by using 
conformal field theory techniques in the dimensionally reduced (1+1)
case. Such an analysis shows that the time dependent geometry around a
collapsing shell will lead to emission of a flux of particles to
infinity though the spectrum, in general, will not be thermal. This
result has nothing to do with black holes or event horizons.   

When the shell does collapse to a radius close to the Schwarzschild
radius, the relevant functions acquire the well known universal form 
characteristic of an exponential redshift of the outgoing 
modes. This exponential redshift leads to a thermal spectrum. In other
words while any collapsing shell will lead to a radiation flux, such a
flux will not, in general, have any universal characteristic. But if
the collapsing shell goes close to its Schwarzschild radius, the flux 
becomes approximately thermal. 

If the collapsing shell follows a trajectory which does not form an
event horizon and --- instead --- 
asymptotes to a radius $2M(1-\epsilon^2)^{-1}$, then we still get
(approximately) thermal radiation during the period $M \lesssim t
\lesssim M\ln (1/\epsilon^2)$.  Of course, in the case $\epsilon=0$,
the collapse leads to an event horizon and we get the standard 
result that thermal radiation will be emitted at all late times
corresponding to $t\gtrsim M$.  On the other hand, for $\eps\neq0$, at
$t>M\ln (1/\epsilon^2)$ the radiation decays to zero exponentially as
to be expected around a final static configuration. 

The above description does not take into account the effect of
backreaction on the formation of event horizon. If radiation of energy
can decrease the mass (and hence the effective radius of the event
horizon) at a sufficiently rapid pace, then the collapsing shell might
never  catch up with the event horizon and a black hole may never
form.  This, however, does not happen essentially because  the amount
of radiation emitted during the  collapse is not enough to
decrease the effective radius of the event horizon sufficiently  fast.  
This result is nontrivial and requires the careful
computations which we have performed in this paper.

\acknowledgments
\noindent 
We thank  T. P. Singh and Dawood Kothawala for helpful
discussions. TP gratefully acknowledges hospitality at TIFR during a
visit in which this work was begun. AP gratefully acknowledges
hospitality at IUCAA during a visit in which a part of this work was
completed. 

\appendix
\section{VEV of stress tensor and related issues}
\noindent 
In this appendix we present a pedagogical review of the calculation of
the VEV of the stress tensor, using the tools of 1+1 conformal field
theory, and highlight certain conceptual issues. The reader is
referred to \Cites{brout,bir-dav,davies,unruh} for details of the
original calculations. The notation used here was laid out in
\Sec{hawkingbasics}. 
\subsection{Choice of vacuum state}
\noindent
In the Heisenberg picture, there is only one vacuum state that one can
work with, which is the \state{in} state. Nevertheless, it is common
to find references in the literature to vacua defined with respect to
various combinations of ingoing and outgoing null coordinates (for a
review of these vacuum states see \Cite{TPreview}). One
particular vacuum known as the Unruh vacuum takes on a special
significance in the collapse situation. This is the vacuum defined to
be positive frequency with respect to the ingoing
Eddington-Finkelstein coordinate $v$ and the outgoing \emph{Kruskal}
coordinate $\Cal{U}=-4Me^{-u/4M}$ \cite{unruh}. As the surface of the
object approaches its Schwarzschild radius, one finds that under the
generic assumption that the trajectory in the $(v,r)$ coordinates
remains well-behaved at the Schwarzschild radius, we get $G(u)\to
-4Me^{-u/4M} = \Cal{U}$ as $u\to\infty$. This behaviour arises due to
the presence of an ingoing ray labelled by $v=v_H$ say, which after
reflection is the \emph{last ray} that can exit the object before the
horizon forms (i.e. before the trajectory crosses $r=2M$). The
presence of this ``last escaper'' ingoing ray means that in the 
 large $u$ limit for the outgoing rays, the corresponding ingoing rays
bunch together around $v\lesssim v_H$, and hence the function $v_{\rm
  ent}(V)$ can be linearized. Additionally, assuming that the
trajectory crosses the Schwarzschild radius at finite $v=v_0$ with a
finite non-zero velocity  $dR_s/dv|_{v_0}$, it is straightforward to
show using \eqns{2eq4} that as $u\to\infty$ (i.e. $R_s\to2M$),
$U(u)\sim e^{-u/4M}+\Cal{O}(e^{-u/2M})$. This shows 
that at late times $G(u)\propto e^{-u/4M}$, and in this limit, the
$G$-vacuum is identical to the Unruh vacuum. 

The previous arguments provide a dynamical ``reason'' for
``choosing'' the Unruh vacuum at late times : the reason is simply the
Heisenberg picture statement that the state does not evolve. The rest
is taken care of by the dynamics and boundary conditions. 

\subsection{Stress tensor VEVs : Regularization}
\noindent
The object of interest is \vev{in}{T_{ab}}, where $T_{ab}$ is the
\emph{2-dimensional} stress tensor in the $u-v$ plane. Provided the
angular components of the stress tensor vanish, the four dimensional
stress tensor can be obtained simply by rescaling the 2-d one by
$(4\pi r^2)^{-1}$. The classical value of $T_{ab}$ for a 
2-d massless scalar field is
$\p_a\vphi\p_b\vphi-(1/2)g_{ab}\p^i\vphi\p_i\vphi$, and it is then  
easy to see by expanding the field operator in the modes $\vphi^{\rm
  out}_\lam$, that \vev{in}{T_{ab}}\ corresponds ultimately to
expectation values of the number operators $b^\dagger_\lam b_\lam$ in
the $G$-vacuum. Operationally one works with an expansion in the
$\vphi^{\rm in}_\om$ modes since the state \state{in} is the vacuum
of the $G$-modes. 

However, the stress tensor VEV needs to be regularized, since even in 
Minkowski spacetime with $\sqrt{4\pi\om}\vphi_\om=e^{-i\om v}-e^{-i\om
  u}$, this object diverges. So what we want is
\be
\avg{T_{ab}}^{\rm ren} = \vev{in}{T_{ab}} - {\rm ``Minkowski\,\,
  value"}\,. 
\la{2eq21}
\ee
%
\subsubsection{The accelerating mirror}
\noindent
Let us first consider the toy example of a scalar field quantized in
flat spacetime ($ds^2 = -dudv + ...$) with the boundary condition
provided by an accelerating mirror which follows the trajectory
$v=G(u)$ for some $G(u)$. Note that since there is no ``interior''
region in this problem, the complication of finding a function like
$v_{\rm ent}(V)$ does not exist. The analysis is therefore simpler,
and will help to set up some
formalism which can then be easily adapted to the case of a genuine
collapse situation. In particular one finds that the renormalized VEV
required can ultimately be obtained in a fairly straightforward
manner, without going into details of contour integrals etc. While the
interesting case occurs when $G(u)=-(1/a)e^{-au}$ for constant $a$, in
which case reflection from the mirror leads to a thermal spectrum of
particles in the $G$-vacuum, we will not require this form of $G$ in
the calculation below. The flat spacetime mode functions satisfying
the reflection condition on $v=G(u)$ are, 
\be
f_\om = \frac{1}{\sqrt{4\pi\om}} \left(e^{-i\om v} - e^{-i\om G(u)}
\right) \,.
\la{2eq22}
\ee
The ``Minkowski value'' for the stress tensor component $T_{vv}$ is
simply \vev{in}{T_{vv}}, while for $T_{uu}$ it is
\vev{out}{T_{uu}}. This follows from the fact that the Minkowski
vacuum has modes which are positive frequency with respect to $u$ and
$v$. This gives us
\be
\avg{T_{vv}}^{\rm ren} = 0 ~~;~~ \avg{T_{uu}}^{\rm ren} =
\frac{1}{12\pi}(G^\prime)^{1/2}\p^2_u(G^\prime)^{-1/2}\,,
\la{2eq23}
\ee
where $G^\prime\equiv dG/du$. When $G^\prime=e^{-au}$, the second
expression can be obtained by explicitly writing out the mode
expansions and performing the integrals involved (see \Cite{brout},
Sec. 2.5.2). In general, the following regularisation procedure proves
to be very useful, since it carries over to the gravitational case as
well (see \Cite{bir-dav} for alternative regularisation procedures).  

One notes that the $u$-part of the Green function for the field in
the $u$-vacuum \state{out}, is $(-1/4\pi)\ln|u-u'|$ where one is
considering two spacetime events $(u,v)$ and $(u',v')$. Similarly, the
Green function in the $G$-vacuum \state{in}\ is
$(-1/4\pi)\ln|G(u)-G(u')|$. The stress tensor VEV in the state
\state{in}\ can then be computed using a point split double
derivative, as 
\begin{align}
\vev{in}{T_{uu}} &= \lim_{u\to u'} \vev{in}{\p_u\vphi(u)
  \p_{u'}\vphi(u')} \nonumber\\
&= \lim_{u\to u'} \p_u\p_{u'}\vev{in}{\vphi(u)\vphi(u')} \nonumber\\ 
&= \lim_{u\to u'} \p_u\p_{u'}\left(-\frac{1}{4\pi}\ln|G(u)-G(u')|
\right)\,, 
\la{2eq24}
\end{align}
and similarly \vev{out}{T_{uu}}, which gives the result in
\eqn{2eq23} after expanding the expressions to third order in
$(u-u')$. In order to get the result in
a form which is generalizable to curved spacetime as well, one notes
that the above argument for calculating the $uu$ components can be
inverted to get the $GG$ components. One would do this by taking a
point split double derivative with respect to $G$ instead of $u$, and
get 
\begin{align}
\avg{T_{GG}}^{\rm ren} = \vev{in}{T_{GG}} - \vev{out}{T_{GG}} &=
-\frac{1}{4\pi} \lim_{G\to G'} 
\p_G\p_{G'} \left( \ln|G-G'| - \ln|u(G)-u(G')|\right) \nonumber\\
&= -\frac{1}{12\pi} \left(\frac{du}{dG}\right)^{1/2}\p^2_G
\left(\frac{du}{dG}\right)^{-1/2} \,.
\la{2eq25}
\end{align}
It is important to note that the $u$ coordinate here 
corresponds to the \emph{Minkowski} $u$.

\subsubsection{Genuine gravitational fields}
\noindent
Let us now turn to the original problem of a body collapsing to form a
black hole. Due to mathematical similarities with the moving mirror
problem, one can use similar ideas here as well, with some subtle
differences \cite{davies,brout}. We still want a regularized version
of \vev{in}{T_{ab}}, the 2-dimensional stress tensor VEV. But now,
there is no globally defined ``Minkowski value'' which one can
subtract. Instead, one subtracts a \emph{locally defined} object
corresponding to the stress tensor VEV in a \emph{locally inertial
  vacuum}. Given any spacetime with  
\be
ds^2=-C(G,v)dGdv + ... \,,
\la{2eq27}
\ee
such that we want expectation values with respect to the vacuum of the
modes $e^{-i\om v}$ and $e^{-i\om G}$, one can show that locally
inertial coordinates are given, at some event $x\equiv(G_0,v_0)$, by  
\be
\hat u = \int_{G_0}^G{\frac{C(G',v_0)}{C(G_0,v_0)}dG'} ~~;~~ \hat v = 
\int_{v_0}^v{\frac{C(G_0,v')}{C(G_0,v_0)}dv'}\,.
\la{2eq28}
\ee
$\hat u$ and $\hat v$ are affine parameters along the radial null
geodesics $v=v_0$ and $G=G_0$ respectively, which pass through $x$,
and the Christoffels in the $(\hat u,\hat v)$ coordinates vanish at
$x$ \cite{brout}. At this event $x$ one wants to subtract from
\vev{in}{T_{ab}}, the quantity \vev{I(x)}{T_{ab}}\ where \state{I(x)}
is the vacuum corresponding to the ``locally inertial modes''
$e^{-i\om\hat u}$, $e^{-i\om\hat v}$ at $x$.

The mathematics is now identical to the moving mirror case, and one
finds
\begin{align}
\avg{T_{GG}}^{\rm ren}_{G} &= -\frac{1}{12\pi}
\left(\frac{d\hat u}{dG}\right)^{1/2}\p^2_G
\left(\frac{d\hat u}{dG}\right)^{-1/2} \,,\nonumber\\
\avg{T_{vv}}^{\rm ren}_{G} &= -\frac{1}{12\pi}
\left(\frac{d\hat v}{dv}\right)^{1/2}\p^2_G
\left(\frac{d\hat v}{dv}\right)^{-1/2} \,,
\la{2eq29}
\end{align}
where the additional subscript on the VEVs reminds us that we are
working in the $G$-vacuum (which is of course also the
$v$-vacuum). \eqn{2eq28} can be used to remove all explicit reference
to the locally inertial coordinates, and we find in general that
\be
\avg{T_{GG}}^{\rm ren}_G = -\frac{1}{12\pi}C^{1/2}\p^2_GC^{-1/2} ~~;~~
\avg{T_{vv}}^{\rm ren}_G = -\frac{1}{12\pi}C^{1/2}\p^2_vC^{-1/2} \,.
\la{2eq30}
\ee
For our case, the conformal factor is given in the exterior
Schwarzschild geometry, by
\be
C = \left(1 - \frac{2M}{r} \right) \frac{du}{dG} \,,
\la{2eq31}
\ee
which follows from comparing the metrics in \eqns{2eq1} and
\eqref{2eq27}, where $u$ is the outgoing Eddington-Finkelstein
coordinate. This simple prescription for constructing the stress
tensor VEV is a consequence of considering only the $s$-wave
contribution and further dropping the Schwarzschild potential barrier,
which is what allowed us to approximate the scalar field modes in the
exterior as propagating on a conformally flat geometry. See
\Cite{brout} for a discussion on how good this approximation is in
practice. 
\subsection{Changing vacua}
\noindent
Let us see how the renormalized stress tensor VEV behaves under a
change of the reference vacuum. This will allow us to write
expressions which will clarify what happens in various physical
situations. The structure of this object is, e.g.
\begin{align}
\avg{T_{GG}}^{\rm ren}_G \sim C^{1/2}\p^2_GC^{-1/2} &\sim~
\underbrace{\p_G\p_{G'}} ~~
\left[~\ln\underbrace{|G-G'|} - \ln|\hat u - \hat u'| ~\right]
\la{2eq32}\\  
&\text{\scriptsize Choice of coords}~~~~\text{\scriptsize Choice of 
  vacuum}
\nonumber
\end{align}
This shows that in any given vacuum, changing coordinates is trivial,
one simply multiplies by the appropriate coordinate transformation
factors. On the other hand, in a given set of coordinates, changing
the \emph{vacuum} is not as trivial, although it is
straightforward. Let us suppose that we want to compute \avg{T_{ab}}\ in the vacuum
corresponding to some null coordinates $(f,g)$ instead of $(G,v)$,
where we assume that $f=f(G)$, $g=g(v)$ so that the 2-d metric remains 
conformally flat. [Later we will specialize to the case $f=u$, $g=v$.]
Our previous results (\eqn{2eq30}) tell us that 
\be
\avg{T_{ff}}^{ren}_{fg} = -\frac{1}{12\pi}
(C_{fg})^{1/2}\p^2_f(C_{fg})^{-1/2}  ~~;~~ \avg{T_{gg}}^{ren}_{fg} =
-\frac{1}{12\pi} (C_{fg})^{1/2}\p^2_g(C_{fg})^{-1/2}\,,
\la{2eq33}
\ee
with the subscript $\avg{}_{fg}$ on the VEVs denoting the choice of
vacuum, and where
\be
C_{fg} = C\left(G(f),v(g)\right)\frac{dG}{df}(f)\frac{dv}{dg}(g)\,.
\la{2eq34}
\ee
In general for some function $F(x)$, we have the identity 
\be
F^{1/2}\p^2_xF^{-1/2} = -\frac{1}{4}\left[\,2\p^2_x\ln F - (\p_x\ln F)^2\,
\right] \,.
\la{2eq35-app}
\ee
Using this and the expression for $C_{fg}$ in \eqn{2eq34}, one can
simplify \eqn{2eq33} to obtain 
\begin{align}
\avg{T_{ff}}^{\rm ren}_{fg}&=\, \left(\frac{dG}{df}\right)^2
\avg{T_{GG}}^{\rm ren}_G - \frac{1}{12\pi}
\left(\frac{dG}{df}\right)^{1/2} \p^2_f
\left(\frac{dG}{df}\right)^{-1/2} \,,\nonumber\\
\avg{T_{gg}}^{\rm ren}_{fg}&=\, \left(\frac{dv}{dg}\right)^2
\avg{T_{vv}}^{\rm ren}_G - \frac{1}{12\pi}
\left(\frac{dv}{dg}\right)^{1/2} \p^2_g
\left(\frac{dv}{dg}\right)^{-1/2}
\la{2eq36}
\end{align}
In particular, for the ``Boulware'' vacuum defined through the
modes $e^{-i\om u}$ and $e^{-i\om v}$, we have $f=u$, $g=v$. The
Boulware vacuum attains physical relevance only in the static case
when the star/shell does not evolve, and one has $G(u)=u+\,$const
(which can be shown using the trajectory and matching equations using 
$R_s=\,$const). But the vacuum itself can of course always be defined
even in the dynamical situation. For this vacuum, 
\begin{align}
\avg{T_{uu}}^{\rm ren}_{B}&=\, 
\avg{T_{uu}}^{\rm ren}_G - \frac{1}{12\pi}
\left(\frac{dG}{du}\right)^{1/2} \p^2_u
\left(\frac{dG}{du}\right)^{-1/2} \,,\nonumber\\
&\nonumber\\
\avg{T_{vv}}^{\rm ren}_{B}&=\, \avg{T_{vv}}^{\rm ren}_G \,.
\la{2eq37}
\end{align}
As discussed earlier, in the limit when the object approaches $r=2M$,
we have $G(u)\to-4Me^{-u/4M}$, and it is not hard to show that
\be
\avg{T_{uu}}^{\rm ren}_B = \avg{T_{uu}}^{\rm ren}_{\Cal{U}} -
\frac{\pi}{12}T_H^2 \,,
\la{2eq39}
\ee
where $T_H = 1/(8\pi M)$ is the Hawking temperature of the collapsing
object, and we have denoted the asymptotic $G$-vacuum by the subscript
$\Cal{U}$ (for $\Cal{U}$nruh). Using the fact that the conformal
factor for the Boulware vacuum is simply $(1-2M/r)$, where $r$ is
given implicitly in terms of $u$ and $v$ by the relation \eqref{2eq2},
the stress tensor VEV in the Boulware vacuum can be explicitly
computed as 
\be
\avg{T_{uu}}^{\rm ren}_B = -\frac{1}{48\pi}\frac{M}{r^3} \left(2 -
\frac{3M}{r} \right)\,.
\la{2eq40}
\ee
This gives us the relations
\begin{align}
\avg{T_{uu}}^{\rm ren}_B(r) &= -\frac{\pi}{12}T^2_H \left[
  \frac{32M^3}{r^3} - \frac{48M^4}{r^4} 
  \right] =\avg{T_{vv}}^{\rm ren}_B = \avg{T_{vv}}^{\rm ren}_{\Cal{U}}
\,, 
\la{2eq41}\\
&\nonumber\\
\avg{T_{uu}}^{\rm ren}_{\Cal{U}}(r) \,&=\, \avg{T_{uu}}^{\rm ren}_B(r)
- \avg{T_{uu}}^{\rm ren}_B(r=2M) \nonumber\\
&= \frac{\pi}{12}T^2_H\left[ 1 -
  \frac{32M^3}{r^3} + \frac{48M^4}{r^4}\right]
\nonumber\\
&= \frac{\pi}{12}T^2_H \left(1-\frac{2M}{r}\right)^2
\left[1+\frac{4M}{r} + \frac{12M^2}{r^2} \right]\,,
\la{2eq42}
\end{align}
where $\avg{T_{uu}}^{\rm ren}_B=\avg{T_{vv}}^{\rm ren}_B$ follows from
the fact that $\p_ur=-\p_vr$. We see that $\avg{T_{uu}}^{\rm
  ren}_{\Cal{U}}$ vanishes on $r=2M$ but attains the constant Hawking
flux value as $r\to\infty$. Also, since the coordinate transformation
factor relevant to the $u$ sector is $d\Cal{U}/du = (-\Cal{U}/4M)$
which goes to zero in the asymptotic limit, we see that the component 
$\avg{T_{\Cal{U}\Cal{U}}}^{\rm ren}_B(r)\to\infty$ in this asymptotic
limit, for \emph{any finite value} of $r$. [The limit $\Cal{U}\to0$ is
  distinct from the limits $r\to2M$ or $r\to\infty$.]

\begin{table}[t]
\centering
\small
\begin{tabular}{|c|c|c|c|}\hline
 & & & \\ [-2ex]
 & $G\to \Cal{U}\to0$; $r=\,${\small const.}$\,<\infty$ & $G\to
  \Cal{U}\to0$; $r\to\infty$ & $G\to \Cal{U}\to0$; $(r-2M)\propto
  \Cal{U}$ \\    
 & {\small Static asymptotic} & {\small Static asymptotic
    observer} & {\small Freely falling} \\ [-0.5ex]
 & {\small observer at finite $r>2M$} & {\small at spatial infinity} & 
  {\small asymptotic observer} \\ [1ex] \hline 
 & & & \\ [-2ex]
$~\avg{T_{uu}}^{\rm ren}_B~$ & \eqn{2eq41} & 0 & $-(\pi/12)T^2_H$
  \\ [1ex]\hline 
 & & & \\ [-2.5ex]
$~\avg{T_{\Cal{U}\Cal{U}}}^{\rm ren}_B~$ & $\infty$ & $0$ {\tiny (if
    $r\to\infty$ first)} &  $-\infty$ \\ [0.5ex]\hline 
 & {\large $\bm{\ast}$} &{\large $\bm{\ast}$} & \\ [-1ex]
$~\avg{T_{uu}}^{\rm ren}_{\Cal{U}}~$ &  \eqn{2eq42} &
  $(\pi/12)T^2_H$ & 0 \\ [0.5ex]\hline 
 & & & {\large $\bm{\ast}$} \\ [-1ex]
$~\avg{T_{\Cal{U}\Cal{U}}}^{\rm ren}_{\Cal{U}}~$ & $\infty$ & $\infty$
  &  Finite, obs. dep. \\ [0.5ex]\hline 
\end{tabular}
\caption{\small Various limiting cases of the stress tensor VEV in
  different vacua.}  
\label{tab1}
\end{table}
The object $\avg{T_{\Cal{U}\Cal{U}}}^{\rm ren}_{\Cal{U}}$ is
trickier. This object has a factor $(4M/\Cal{U})^2$ \emph{and} a
factor $(1-2M/r)^2$ in the asymptotic limit. Clearly it remains finite
\emph{on the horizon} if one takes the $\Cal{U}\to0$ and $r\to2M$
limits simultaneously, assuming $\Cal{U}\propto(r-2M)$. Such a
situation is relevant for an \emph{asymptotic freely falling observer}
maintaining approximately constant $v$, for example a
geodesic observer who starts from rest at some very large distance.
This result therefore shows that such an observer must see a finite
flux at late times, although apparently this flux is not universal and
will depend on the specific trajectory of the freely falling
observer. 

For $r=\,{\rm const.\,} > 2M$, the Kruskal coordinates are not the
natural choice; such an observer uses the Eddington-Finkelstein
$(u,v)$ null coordinates. Hence, even though
$\avg{T_{\Cal{U}\Cal{U}}}^{\rm ren}_{\Cal{U}}$ diverges as
$\Cal{U}\to0$ for fixed $r$, this is not physically relevant. The
natural choice  for such an observer in the $\Cal{U}\to0$ limit is
$\avg{T_{uu}}^{\rm ren}_{\Cal{U}}(r)$, which as we see remains finite
and attains the Hawking flux value at large distances
$r\to\infty$. \tab{tab1} summarizes these results. The boxes
highlighted by asterisks show physically relevant limits. Not
surprisingly, all of these are finite. 
\subsection{Stress tensor VEV at arbitrary times}  
\noindent
The calculations in the previous section show that in the limit when
the collapsing object approaches its Schwarzschild radius, the stress
tensor VEV becomes time independent and takes the form given in
\eqns{2eq42} and \eqref{2eq41} for the components $\avg{T_{uu}}^{\rm
  ren}_{G}$ and $\avg{T_{vv}}^{\rm ren}_{G}$ respectively, in the
limit $G\to\Cal{U}$. Physically this corresponds to a steady blackbody
flux of particles at any radius $r$ given by $T^r_t = T_{vv} - T_{uu}
= -(\pi/12)T_H^2$. Showing that this flux is associated with a thermal 
spectrum requires a calculation of the Bogolubov coefficients relating
the Boulware and Unruh vacua (see e.g. \Cite{brout}). We will not
discuss this calculation since we wish to discuss backreaction, for
which we need the flux and not the explicit form of the particle
spectrum.   
In principle though, one already has the form of the stress tensor VEV
at \emph{any} stage during the collapse, since we have derived the
relation (see \eqns{2eq37} and \eqref{2eq41})
\begin{align}
\avg{T_{uu}}^{\rm ren}_G - \avg{T_{vv}}^{\rm ren}_G(r) = \frac{1}{12\pi}
\left(\frac{dG}{du}\right)^{1/2} \p^2_u
\left(\frac{dG}{du}\right)^{-1/2} \equiv \avg{T_{uu}}^{\rm traj}_G(u)\,,
\la{2eq43-app}
\end{align}
which defines the trajectory-dependent quantity $\avg{T_{uu}}^{\rm
  traj}_G(u)$ which asymptotes to $(\pi/12)T_H^2$ at late stages of
the collapse. Here $\avg{T_{vv}}^{\rm ren}_G(r)=\avg{T_{uu}}^{\rm
  ren}_B(r)$ is given by \eqn{2eq40} or \eqn{2eq41}.
\section{Proofs for various results}
\noindent
In this appendix we give calculational details for some of the results
quoted in the main text.
\subsection{Asymptotic behaviour of the trajectory defined by
  \eqn{2eq45}} 
\label{appendix1}
\noindent
As mentioned in the text, for $U^\prime$ given by \eqn{2eq45}, the
trajectory equation \eqref{2eq48} forces $(1-2M/R_s)$ to approach
$\eps^2$ at late times. Let us assume therefore, that for large enough 
times the asymptotic behaviour is given by
\be
1-\frac{2M}{R_s} = \eps^2 + f(u) ~~;~~ |f(u)|\ll\eps^2\,.
\la{app1}
\ee 
We will soon determine the condition to be satisfied by $u$ in order
for this to hold. Differentiating and rewriting \eqn{app1} leads to
\be
2R_s^\prime = \left(\frac{R_s}{2M}\right)^24Mf^\prime ~~;~~
\frac{R_s}{2M} = \frac{1}{1-\eps^2-f}\,,
\la{app2}
\ee 
which are exact. The behaviour of $U^\prime$ in the asymptotic regime
is well approximated by $U^\prime = \eps^2+\ti{A}e^{-u/4M}$, where
$\ti{A}$ is different from the $A$ which appears in \eqn{2eq45}. The
assumption that the asymptotic behaviour for the function $h(u)$ in
\eqn{2eq47} is achieved exponentially on a timescale $\kappa^{-1}$,
means that we can expect $\ti{A}=\Cal{O}(e^{1/\kappa M})$. Recall from
the main text that we have 
\be
\kappa<\kappa_{\rm max}\sim M^{-1}(M/R_0)^b ~~;~~b>0\,.
\la{app-extra1}
\ee
Using the expressions in \eqn{app2} in the trajectory equation
\eqref{2eq48} and linearising in $f$ leads to
\be
4Mf^\prime(1-\eps) \approx (1-\eps^2)^2\left(2\eps\ti{A}e^{-u/4M} - f
\right)\,. 
\la{app3}
\ee 
Defining the variable $s\equiv u/4M$, we have
\be
\p_s f + K_1f = K_2e^{-s}\,,
\la{app4}
\ee 
where
\be
K_1\equiv \frac{(1-\eps^2)^2}{(1-\eps)} ~~;~~ K_2 = 2\eps\ti{A}K_1\,. 
\la{app5}
\ee 
Solving \eqn{app4} gives
\be
f(u) = C e^{-K_1u/4M} + \frac{K_2}{K_1-1}e^{-u/4M} ~~;~~ {\rm if~}
K_1\neq1\,, 
\la{app6}
\ee 
with $f=(C+K_2u/4M)e^{-u/4M}$ when $K_1=1$, which we ignore since this
solution forms a set of measure zero. The condition $|f|\ll\eps^2$
shows that for consistency we must have
\be
u\gg\kappa^{-1}\ln(1/\eps^2)\,,
\la{app6-extra1}
\ee 
which follows from the fact that for small $\eps$,
$K_2/(K_1-1)\sim\ti{A}\sim e^{1/\kappa M}$. The constant $C$ is not 
constrained by this analysis : it can only be fixed by evolving the
full set of equations. The best we can say is that it must be at most
of order $\sim K_2/(K_1-1)$, to be consistent with the condition
$|f|\ll\eps^2$. Consequently the sign of $f$ is also not fixed by
this analysis. We see that $f(u)$ exponentially decays and hence
consistently reproduces $(1-2M/R_s)\to\eps^2$ at late times. 
\subsection{Behaviour of the $U$-dependent backreaction and bound on
  mass loss} 
\label{appendix6}
\noindent
In this section we discuss some features of the backreaction term
$(U^\prime)^{1/2}\p_u^2(U^\prime)^{-1/2}$, and derive an upper bound
on the expected mass loss due to backreaction in the ``asymptotic''
trajectory of \Sec{timelike}.
\subsubsection{Analysing \fig{backU}}
\noindent
An important feature in \fig{backU} is the time
($u=u_\ast$ say) at which any given curve starts deviating from the
$\eps=0$ case. This time depends not only on $\eps$ but also on
$\kappa$. (For now we ignore the dependence on the starting radius
$R_0$ and assume $\kappa<\kappa_{\rm max}$.) We see that for a given
$\kappa$, smaller values of $\eps$ will lead to a larger $u_\ast/M$,
which can be understood as the fact that it will take longer for a
trajectory to ``feel the effect'' of a smaller $\eps$. 

On the other
hand, for a given $\eps$, changing the value of $\kappa$ has three
distinct effects. 
\begin{itemize}
\item Firstly, reducing $\kappa$ ``stretches out'' the
\emph{standard} $\eps=0$ behaviour of the backreaction, which is
expected since the timescale over which the standard trajectory
approaches $r=2M$ is governed by $\kappa^{-1}$ which is now
larger. 
\item Secondly, reducing $\kappa$ increases the value of $u_\ast/M$
at which the backreaction deviates from the standard case, which is
also expected since a larger $\kappa^{-1}$ will slow down the rate at
which $\alpha(u)$ increases and hence delay the time at which the
effects of $\eps\neq0$ are felt. 
\item The third effect however is somewhat
subtle and will play an important role below in determining a bound on
the mass loss due to backreaction for such a trajectory. This effect
is the fact that the increase in $u_\ast/M$ is \emph{generically less}
than the increase in $\kappa^{-1}/M$. In other words, for two values
$\kappa_2<\kappa_1$, although we will have $u_{\ast2}>u_{\ast1}$, we
also see that $\kappa_2u_{\ast2}<\kappa_1u_{\ast1}$. Indeed,
\fig{backUa} shows that for the chosen values of $\kappa$,
$\kappa_1u_{\ast1}>1$ whereas $\kappa_2u_{\ast2}\simeq1$. To
understand the reason behind this behaviour, we note that the value of
$u_{\ast}$ for some $\kappa$ and $\eps$ is essentially determined by
the condition $e^{\alpha(u_\ast)}\sim A/\eps$ which follows from the
structure of the right hand side in \eqn{2eq50}. Also, $\alpha(u)$ is an integral
of a function  $h(u)$ which in turn is controlled by $\kappa$. If we
reduce $\kappa$, then even though $h(u)$ takes a longer time to rise
to its maximum value of $1$, the integral of $h(u)$ can pick up a
sizeable portion of its required budget of order $\sim M\ln(A/\eps)$
even for $u<\kappa_2$. This makes it plausible that although 
$u_{\ast2}>u_{\ast1}$, we in fact obtain
$\kappa_2u_{\ast2}<\kappa_1u_{\ast1}$. It is difficult to give a more
precise analytical argument for this observed fact.
\end{itemize}       
\subsubsection{A bound on mass loss}
\noindent
The upshot is that reducing $\kappa$ also reduces the maximum
magnitude of the $U$-dependent backreaction term. We can use this fact
to construct an upper bound on the mass loss due to backreaction along
this ``asymptotic'' trajectory. Since the $V$-dependent term remains
subdominant until $u_{\rm ent}\sim\kappa^{-1}$, and thereafter is
expected to follow a profile qualitatively similar to the
$U$-dependent term with slightly different timescales, for order of
magnitude estimates we will not consider this term separately, and
instead pretend that the $U$-dependent term itself accounts for all
the backreaction for the entire duration starting from $u=0$. Our
estimate for mass loss is a simple one : we ask for the maximum
possible backreaction over the maximum possible time, and take the
product. This will serve as a proxy for the integral
$\int{\avg{T_{uu}}^{\rm traj}du}$ (which is the mass loss at leading
order, see \Sec{bkrxn}).  

To begin with, we notice that for any value of $\kappa$, the mass loss
will be enhanced for smaller values of $\eps$, since reducing $\eps$
will in principle allow the backreaction to be sustained at its
maximum value of $\sim M^{-2}$ (although the value of $\kappa$ may
prevent this). This is clear from \fig{backUa}. So for any $\kappa$,
the largest mass loss will be achieved when $\eps$ saturates its
Planck scale bound (see main text, \Sec{bound}) of $\eps^2\geq1/2M$. 

Having fixed $\eps$, we see that if we choose $\kappa$ appropriately
so that the backreaction remains at its maximum value for times
$\kappa^{-1}\lesssim u\lesssim\kappa^{-1}\ln(1/\eps^2)$, then the mass
loss is approximated by $\Delta M\sim
M^{-2}\kappa^{-1}\ln(1/\eps^2)$. The question now is what value of
$\kappa$ will maximize this quantity. Naively one might think that
making $\kappa$ \emph{smaller} will help the case, since reducing
$\kappa$ increases the time over which the backreaction is
in effect. But as we saw in the previous subsection, this is not the
whole story and there is a competing effect at play : reducing
$\kappa$ will also mean reducing the peak magnitude of the
backreaction, thus decreasing the mass loss. Clearly there is an
optimum value for $\kappa$ which will maximize the mass
loss. \fig{backUb} and similar figures obtained by varying $\kappa$ for
fixed $\eps$ suggest that for values of $R_0$ significantly larger
than $2M$, this optimum value is in fact the
\emph{maximum} value $\kappa_{\rm max}$, which is also borne out by
numerically integrating the $U$-dependent term for various values of
$\kappa$ (and $\eps$ as well). 

Finally, we note that $\kappa_{\rm max}$ is a decreasing function of
the starting radius $R_0$, which we estimated earlier as $\kappa_{\rm
  max}\sim M^{-1}(M/R_0)^b$ for $R_0$ significantly larger than $2M$,
where $b>0$ in general and $b\approx2$ for
$h(u)=\theta(u)\tanh(\kappa^2u^2)$. If this scaling were true for all
values of $R_0$, then a simple estimate for the
largest possible value of the backreaction would come from setting
$\kappa=\kappa_{\rm max}$ with $R_0/2M\gtrsim\Cal{O}(1)$ (for an
$\eps$ which saturates the Planck scale bound), i.e. by setting
$\kappa\sim M^{-1}$ up to numerical factors. However this scaling
does not hold when $R_0$ is close to $2M$ and $\kappa_{\rm max}$ in
this case can actually be very large. Physically this is because it is
possible for the trajectory to start its asymptotic phase very quickly
without becoming superluminal. It might then seem that asymptotic
trajectories which start close to $r=2M$ and very quickly enter their
asymptotic phase will radiate away a negligible amount of mass. This
is not true since the form of the \emph{backreaction} in the initial
stages of collapse also changes in this case : notice that the
$U$-dependent term \eqref{2eq50} depends on
$\alpha^{\prime\prime}=h^\prime/4M$, and for large enough $\kappa$,
$h$ approaches the Heaviside step function so that $h^\prime$ shows a
spike close to $u=0$ of strength $\sim M^{-1}$. The bottomline is that
even for small values of $R_0$ for which $\kappa_{\rm max}$ can be
very large, the mass loss is still of the same order of magnitude as
it would be in our estimate $M^{-2}\kappa^{-1}\ln(1/\eps^2)$, with
$\kappa\sim M^{-1}$ and $\eps^2\sim M^{-1}$. The final upper bound on
the mass loss (ignoring numerical factors) is given by
\be
\Delta M_{\rm max} = \Cal{O}(M^{-1}\ln(M))\,.
\la{masslossbound}
\ee
%
\subsection{Stress tensor VEV in initial phase of trajectory \eqn{2eq45}} 
\label{appendix3}
\noindent
Here we show that for type I rays which enter at times
$0<u\lesssim \kappa^{-1}\ln(1/\eps^2)$, the stress tensor VEV is at
most of order $\Cal{O}(1/M^2)$ and changes on a timescale $\sim M$. The
argument relies on simple order of magnitude estimates using the fact
that the velocity in proper time $\beta\equiv dR_s/d\tau$ in this
phase satisfies $\beta =  \Cal{O}(1)$ as we show below. [$\beta$
  actually starts from zero, rises in magnitude, and then falls as the
  asymptotic phase begins, but for simplicity
  we will treat it as order unity throughout this phase.]
We only need to concentrate on the second term in \eqn{2eq44} since we
already have an exact expression \eqref{2eq50} for the first.

Writing the trajectory matching equations \eqref{2eq4} in terms of the
proper time $d\tau^2=-ds^2|_{\rm traj}$ and noting that $\beta$ is
negative for a collapsing trajectory, we have the following relations 
\be
\frac{dV}{d\tau} = \sqrt{1+\beta^2}+\beta ~~;~~ \frac{dv}{d\tau} =
\frac{1}{(1-2M/R_s)}\left(\sqrt{1 - 2M/R_s+\beta^2}+\beta \right) \,,
\la{app13}
\ee
\be
\frac{dU}{d\tau} = \sqrt{1+\beta^2}-\beta ~~;~~ \frac{du}{d\tau} =
\frac{1}{(1-2M/R_s)}\left(\sqrt{1 - 2M/R_s+\beta^2}-\beta \right) \,.
\la{app14}
\ee
Using these, one can write down expressions for $U^\prime$ and $\dot
V$ in terms of $\beta$ and $R_s$. Further, in the initial phase of the
trajectory we have $U^\prime=\eps+Ae^{-\alpha(u)}=\Cal{O}(1)$,
$(1-2M/R_s)=\Cal{O}(1)$ and hence $dv/du=\Cal{O}(1)$. This implies
that $\beta$ and hence $\dot V$ are of order unity.

The second term in \eqn{2eq44} can therefore be estimated as $\sim
\tau_V^{-2}$ where $\tau_V$ is the timescale controlling $\dot
V$. This timescale is set entirely by the function $U^\prime$ due to
the way we have chosen to define the trajectory. We can therefore
estimate $\tau_V^{-1}$ as
\be
\tau_V^{-1}\sim\p_u\ln U^\prime \sim \alpha^\prime= h(u)/4M\,,
\la{app14-extra1}
\ee 
other factors being of order unity. Due to the required asymptotic
behaviour of $h(u)$, we can approximately take $h(u)\sim0$ for
$u\lesssim\kappa^{-1}$ and $h(u)\sim1$ for $\kappa^{-1}\lesssim
u\lesssim\kappa^{-1}\ln(1/\eps^2)$. This is not too bad an
approximation since, for example, if $h(u)=\tanh(\kappa^2u^2)$, we
will have $h\sim\kappa^2u^2$ for $u\ll\kappa^{-1}$ and
$h\sim1+\Cal{O}(e^{-2\kappa^2u^2})$ for $u\gg\kappa^{-1}$. This
estimate shows that the second term
$(U^\prime/\dot V)^2{\dot V}^{1/2}\p^2_v{\dot V}^{-1/2}$ is expected
to be significant during the interval $\kappa^{-1}\lesssim
u_{\rm ent}\lesssim\kappa^{-1}\ln(1/\eps^2)$ in which its value is at
most $\Cal{O}(1/M^2)$ and changes on a timescale $\sim M$. 
\subsection{Stress tensor VEV in asymptotic phase of trajectory \eqn{2eq45}} 
\label{appendix2}
\noindent
Here we show that the second term in \eqn{2eq44} exponentially decays
for the type II rays which enter and exit the shell in the
asymptotic phase $u\gg\kappa^{-1}\ln(1/\eps^2)$ of the trajectory
\eqn{2eq45}. Using the late time solution \eqref{app6}, we can
parametrically obtain $v_{\rm ent}(V)$ by finding $v(u)|_{\Cal{P}_{\rm
    ent}}$ and $V(u)|_{\Cal{P}_{\rm ent}}$ as follows : From
\eqns{2eq4} we get 
\be
\frac{dv}{du} = 1 + \frac{2R_s^\prime}{1-2M/R_s} = 1
+ \Cal{O}\left(\frac{\ti{A}}{\eps^2}e^{-\Cal{K}u/4M}\right) ~~;~~ 
\dot V = \frac{1}{U^\prime}(1-2M/R_s) = \eps\left( 1 +
\Cal{O}\left(\frac{\ti{A}}{\eps^2}e^{-\Cal{K}u/4M}\right) \right)  \,, 
\la{app7}
\ee 
where $\Cal{K}=\Cal{O}(1)$ is determined from the solution
\eqref{app6} and $\ti{A}=\Cal{O}(e^{1/\kappa M})$ was defined in
Appendix \ref{appendix1} above. This leads to 
\be
v_{\rm ent}(V) = \frac{V}{\eps}  +
\Cal{O}\left(\frac{M\ti{A}}{\eps^2}e^{-\Cal{K}V/4M\eps}\right) \,. 
\la{app8}
\ee 
Further, since the exit point is also in the asymptotic phase, we have
\be
U = \eps \left( u + \Cal{O}\left(\frac{M\ti{A}}{\eps}e^{-u/4M}\right)
\right)\,, 
\la{app9}
\ee
which leads to
\be
v_{\rm ent}(U(u)) = u +
\Cal{O}\left(\frac{M\ti{A}}{\eps^2}e^{-\Cal{K}u/4M}\right) \,.
\la{app10}
\ee 
Using the expression for $\dot V$ from \eqn{app7} above and replacing
$u$ (which is actually $u_{\rm ent}$) by $v$ at the leading order, one
can easily show that 
\be
{\dot V}^{1/2}\p^2_v {\dot V}^{-1/2} =
\Cal{O}\left(\frac{\ti{A}}{\eps^{2}M^2}e^{-\Cal{K}v/4M}\right) \,,
\la{app11}
\ee
and further, evaluation at $v=v_{\rm ent}(U(u))$ simply replaces $v$
by $u$ (which is now $u_{\rm ex}$). Also, at the leading order we
have $U^{\prime2}/{\dot V}^2 \approx 1$, and hence the second term in
\eqn{2eq44} becomes
\be
\frac{U^{\prime2}}{{\dot V}^{2}} \left. {\dot V}^{1/2}\p^2_v {\dot
  V}^{-1/2} \right|_{v=v_{\rm ent}(U(u))} =
\Cal{O}\left(\frac{\ti{A}}{\eps^{2}M^2}e^{-\Cal{K}u/4M}\right) \,,
\la{app12}
\ee
with the approximation valid for $u\gg\kappa^{-1}\ln(1/\eps^2)$.
\subsection{The effect of backreaction on the background geometry} 
\label{appendix4}
\noindent
In this section we sketch the arguments presented by Brout et
al. which show that the late time backreaction is
determined by the time dependent mass $M(u)$ of the collapsing
object. The reader is referred to Sec 3.4 of \Cite{brout} for further
details. The argument proceeds as follows : The spherically symmetric
exterior metric is written in the form
\be
ds^2_{\rm ext} = -e^{2\psi}\left(1-\frac{2m(v,r)}{r} \right)dv^2 +
2e^\psi dvdr + r^2d\Omega^2\,,
\la{app15}
\ee
in terms of which the Einstein equations become
\be
\frac{\p m}{\p v} = T{}^r_v ~~;~~ \frac{\p m}{\p r} = -T{}^v_v ~~;~~
\frac{\p \psi}{\p r} = T{}^r_r/r\,, 
\la{app16}
\ee
where the right hand sides contain the \emph{two-dimensional} stress
tensor components. 

At large distances, the zeroth order calculation shows that only
$T_{uu}=L_H$ survives. (We have already computed $T_{vv}$ which falls
like $\sim(M/r)^3$, and the energy-momentum conservation equation can
be used to show that so does $T_{uv}$.) Hence the metric must
correspond to the outgoing Vaidya solution
\be
ds^2_{{\rm ext, large}\,r} = -\left(1-\frac{2M(u)}{r} \right)du^2 -
2dudr + r^2d\Omega^2 ~~;~~ \frac{dM(u)}{du} = -T_{uu}\,, 
\la{app17}
\ee
where we assume that $L_H=\Cal{O}(1/M^2)$ and varies slowly. Matching
\eqref{app15} and \eqref{app17} at some large radius (say
$r>\Cal{O}(6M)$) gives the transformation between the $(v,r)$ and
$(u,r)$ coordinates as 
\be
e^\psi dv = du + \frac{2dr}{1-2M(u)/r} ~~;~~ m(v,r)=M(u) \,.
\la{app18}
\ee
Using this to calculate the right hand sides of the Einstein equations
\eqref{app16} in terms of $L_H$, one finds that at large $r$ all three
stress tensor components $T{}^r_v$, $T{}^v_v$ and $T{}_{rr}$, are of
order $\Cal{O}(L_H)$ times some quantity of order unity.

Near the apparent horizon $r_{\rm ah}(v) = 2m(v,r_{\rm ah}(v))$, one
uses the fact that $r$ and $v$ both behave like inertial light-like
coordinates (see \eqn{app15}), to argue that $T{}^v_v$ and $T{}_{rr}$
must be of order $\Cal{O}(L_H)$ here as well. Integrating the
conservation equation $T{}^r_{v,r}+T{}^v_{v,v}=0$ from $r_{\rm ah}$ to
$r>\Cal{O}(6M)$ then gives $T{}^r_v=\Cal{O}(L_H)$ near the horizon
also. Specifically, one finds near the horizon (from Eqn. 3.86 of
\Cite{brout})  
\be
\frac{\p m}{\p v} = -L_He^\psi + \Cal{O}(ML_{H,v}) ~~;~~ \frac{\p
  m}{\p r} = \Cal{O}(L_H) ~~;~~ \frac{\p \psi}{\p r} =
\Cal{O}(L_H/r)\,,
\la{app19}
\ee
which shows that the metric function $m$ varies slowly and also that
$\psi$ can be safely set to zero in the calculation.

The next step is to obtain the equation for outgoing null
geodesics (ONGs). This is somewhat involved due to the presence of the 
apparent horizon where the ONGs change the sign of their ``velocity''
$dr/dv$. The analysis proceeds by first noting that as long as $L_H$
is small and slowly varying, the location of the \emph{event horizon}
$r=r_{\rm eh}(v)$ is not very far removed from that of the apparent
horizon. To see this, we write $r_{\rm eh}(v) = r_{\rm ah}(v) +
\Delta(v)$. Differentiating with respect to $v$ and using $2dr_{\rm
  eh}/dv=1-2m(v,r_{\rm eh})/r_{\rm eh}$ (since the event horizon by
definition is the last ONG to reach $\Cal{I}^+$), we find at leading
order in $L_H$ 
\be
\Delta = r_{\rm ah} r_{{\rm ah},v} \sim -1/M \Rightarrow r_{\rm eh} = 
r_{\rm ah}(1+\Cal{O}(L_H))\,,
\la{app20}
\ee
since $dr_{\rm ah}/dv=\Cal{O}(L_H)$. A useful result which follows
from this is $2m(v,r_{\rm eh})=r_{\rm eh}(1+\Cal{O}(L_H))$,
\be
2m(v,r_{\rm eh})=2m(v,r_{\rm ah}+\Delta) =2m(v,r_{\rm
  ah})+\Delta\Cal{O}(L_H) + \ldots = r_{\rm ah}(1+ \Cal{O}(L_H^2)) =
r_{\rm eh}(1+\Cal{O}(L_H))\,. 
\la{app21}
\ee
The ONGs are easier to analyse in terms of the coordinate $x=r-r_{\rm
  eh}(v)$, in terms of which the metric becomes
\be
ds^2_{\rm ext} = -dv^2\frac{2m(v,r_{\rm eh}+x)x}{r_{\rm eh}(r_{\rm
    eh}+x)}(1+\Cal{O}(L_H)) + 2dvdx + r^2d\Omega^2\,. 
\la{app22}
\ee
The coefficient $g_{vv}$ vanishes on the \emph{event horizon} ($x=0$)
in these coordinates rather than the apparent horizon, and the ONGs
near the horizon resemble those in the unperturbed Schwarzschild
geometry, namely we have for $x\ll r_{\rm eh}$,
\be
\ti{v} - 2\ln x = f(u) ~~;~~ \ti{v} = \int^v{\frac{dv}{r_{\rm 
    eh}^2}2m(v,r_{\rm eh})(1+\Cal{O}(L_H))} \,,
\la{app23}
\ee
along an ONG, and the following ansatz describes ONGs at arbitrary
distances $x$,
\be
\ti{v} - 2\frac{x}{r_{\rm eh}(v)} - 2\ln x + \delta =
\int^u{\frac{du'}{\ti{m}(u')}} + D\,. 
\la{app24}
\ee
Here $D$ is a constant of integration, and
$\delta=\Cal{O}(L_H(Mx+x^2)/M)$ which follows from integrating 
the ONG equation $u=\,$const using the form of the metric
\eqref{app22}. By requiring that at large distances the coordinate $u$
be identical to the one appearing in the outgoing Vaidya metric
\eqref{app17}, one finds $\ti{m}(u) = M(u)(1+\Cal{O}(L_H))$ using the
transformation equation \eqref{app18}.

This coordinate transformation is then used to calculate the quantity
$\avg{T_{uu}}^{\rm ren}$ in the same way as was done in the case
without backreaction, and to leading order one finds that the flux on
the horizon vanishes quadratically and the flux reaching $\Cal{I}^+$
is $(\pi/12)T^2_H(u)$ where
\be
T_H(u)\simeq\frac{1}{8\pi \ti{m}(u)} = \frac{1}{8\pi
  M(u)}(1+\Cal{O}(L_H))\,, 
\la{app25}
\ee
which completes the argument (see Eqns. 3.91-3.95 of \Cite{brout} for
details). 
\subsection{Delay in horizon formation for a null trajectory} 
\label{appendix5}
\noindent
In this section we analyse the effect of backreaction on the horizon
formation time for a null in-falling trajectory. Our classical
trajectory in the absence of backreaction has two phases : 
\begin{enumerate}
\item {\bf $u<0$ :} $R_s(u) = R_0 = \,$const.
\item {\bf $u>0$ :} $v=\,$const., $V=\,$const.
\end{enumerate}
The horizon is formed at finite $(U,V)$. The trajectory solution is
\begin{align}
\textbf{Phase (1):~} &U(u) = \left(1-\frac{2M}{R_0}\right)^{1/2}u
  \,-\, 2R_0 + 4M~~;~~ \nonumber\\
&V(v) = \left(1-\frac{2M}{R_0}\right)^{1/2}v
  \,-\, 2x(R_0)\left(1-\frac{2M}{R_0}\right)^{1/2} + 4M ~~;~~ v =
  u+2x(R_0)\,, 
\la{null-eq2}\\
\textbf{Phase (2):~} &V=4M~~;~~ v=2x(R_0)~~;~~ \nonumber\\ 
&U = -2R_s + 4M ~~;~~ u = -2x(R_s) + 2x(R_0)\,, 
\la{null-eq3}
\end{align}
where $x(r)\equiv r +2M\ln(r/2M-1)$ is the ``tortoise'' function and
constants are chosen so that the horizon is formed at $U=0$ (with
$u\to\infty$). The Penrose diagram for this spacetime is shown in
\fig{pen-null}. 
\begin{figure}[t]
\centering
\includegraphics[height=0.35\textheight]{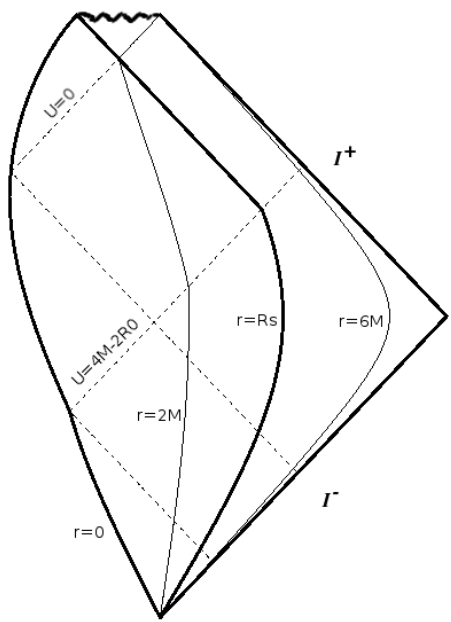}
\caption{Penrose diagram (in the absence of backreaction) for a shell
  which stays at $r=R_0=3.5M$ until $U=4M-2R_0$ and then collapses
  along $v=\,$const., forming a horizon at $U=0$. The shell trajectory
  and the $r=2M$ surface are labelled. The two null rays marking the
  onset of phase (2) and horizon formation, are also shown.}  
\la{pen-null}
\end{figure}
Causality implies that backreaction cannot have any effect on the
spacetime region $U<4M-2R_0$, i.e. while the trajectory is in phase
(1). In phase (2) the trajectory will be modified and will no longer
in general have $V=4M=\,$const. Say the modified trajectory
(parametrized by $U$) is given by $(V=\ti{V}(U),r=\ti{R}_s(U))$ in
interior coordinates and $(v=\ti{v}(U),r=\ti{R}_s(U))$ in exterior
coordinates. The interior geometry is still Minkowski while the
exterior is the Brout et al. approximation
\be
ds^2_{\rm ext} \approx -\left(1-\frac{2m(v,r)}{r}\right)dv^2 + 2dvdr +
r^2d\Omega^2 \,,
\la{null-eq4}
\ee
where the mass function $m(v,r)$ is slowly varying in the entire
exterior. All this refers to phase (2) of the trajectory. The matching
therefore leads to 
\be
d\ti{V} - dU = 2d\ti{R}_s ~~;~~ -dU^2-2dUd\ti{R}_s = -\left(1-
\frac{2m(\ti{v},\ti{R}_s)}{\ti{R}_s} \right)d\ti{v}^2 +
2d\ti{v}d\ti{R}_s \,,
\la{null-eq5}
\ee
and we will assume that the functional form of the trajectory is
shifted by terms of order $\Cal{O}(L_H)$ so that 
\be
\ti{R}_s(U)=R_s(U)(1 + \Cal{O}(L_H)) ~~;~~ R_s(U) =
-\frac{1}{2}(U-4M_0) \,, 
\la{null-eq6}
\ee
with $M_0$ denoting the unperturbed mass of the shell. For the
backreaction $L_H$, \emph{at the leading order} we will take this 
to be given by the exact calculation of \Sec{null} (see
\eqn{toy-eq21}), so that  
\be
L_H = \avg{T_{uu}}^{\rm traj}(R_s) =
\frac{1}{48\pi}\frac{M_0}{R_s^3}\left( 2 - \frac{3M_0}{R_s} \right)\,.
\la{null-eq6-extra1}
\ee
Although the calculation
of \Sec{null} was for a trajectory which does not form a horizon,
the derivation can be easily modified to show that the stress tensor
VEV (in the absence of backreaction) in the present case, will have
the form \eqref{null-eq6-extra1} at least for $R_s>2M_0$. Using this
form for $L_H$ is self-consistent so long as $L_H\ll1$, which is the
case for all $R_s\geq2M_0$. We then have 
\be
1+2\frac{d\ti{R}_s}{dU} = 
\Cal{O}(L_H)\,,
\la{null-eq7}
\ee
which gives us, at the leading order,
\be
\frac{d\ti{V}}{dU} = 
\Cal{O}(L_H) ~~;~~ \frac{d\ti{v}}{dU} = 
\Cal{O}(L_H)  \,.
\la{null-eq8}
\ee
Now assume that the horizon forms at $U=\delta U$ instead of $U=0$, so
that 
\be
\ti{R}_s(\delta U) = r_{\rm eh}(\ti{v}(\delta U)) ~~;~~ \ti{v}(\delta
U) \equiv v_0 + \delta v\,,
\la{null-eq10}
\ee
where $v_0=2x(R_0)$ is the unperturbed value at which the horizon is
formed. We would like to estimate $\delta U$ and $\delta v$. Using
\eqn{null-eq8} we have
\be
\ti{v}(\delta U) = v_0 + \int_{-2R_0+4M_0}^{\delta
  U}{\Cal{O}(L_H)dU}\,, 
\la{null-eq11}
\ee
where the lower limit of integration is set by requiring that the
trajectories in the presence and absence of backreaction be identical
at the beginning of phase (2) which is at $U=-2R_0+4M_0$. At this
level of approximation, using the expression \eqref{null-eq6-extra1}
for $L_H$ with $R_s(U)$ given by \eqn{null-eq6}, we find 
\be
\int_{-2R_0+4M_0}^{\delta U}{L_HdU} =
\left.\frac{M_0}{6\pi}\frac{(2M_0-U)}{(4M_0-U)^3}
\right|_{-2R_0+4M_0}^{\delta U} \,,
\la{null-eq11-extra1}
\ee
and hence, ignoring numerical factors, 
\be
\frac{\delta v}{M} = \Cal{O}(M^{-2})\left(1+ \Cal{O}(\delta U/M) +
\Cal{O}((M/R_0)^2)\right)\,.
\la{null-eq12}
\ee
We can further assume $r_{\rm eh}(\ti{v}(\delta U)) =
2M_0(1+ 
\Cal{O}(M^{-2}))$ since $r_{\rm eh}(v)$ is a slowly
varying function whose unperturbed form is the constant $2M_0$, and
since $L_H=\Cal{O}(M^{-2})$ around horizon formation. Then
\eqn{null-eq10} gives us  
\be
\delta U/M = \Cal{O}(M^{-2}) \,,
\la{null-eq13}
\ee
so that $|\delta U|/M\ll1$ as expected. The signs of $\delta v$ and
$\delta U$ are not determined by this analysis. As in the timelike
case, the backreaction  in this case cannot halt the horizon
formation, but can only delay it by a time  $\Cal{O}(1/M)$.  


\begin{thebibliography}{10}
\bibitem{footnote1}
For completeness, we also note that had we done the analysis with
$\eps=0$ exactly, we would find that as $u\to\infty$, $v_{\rm ent}(V)$
linearises around the ``last escaper'' ray (see Appendix A 1), and the
contribution from the second term is suppressed like $\sim
e^{-u/2M}$. In practice, since we can expect $u_{\rm ex}(u_{\rm
  ent}=0) > 2M$, the second term will remain highly suppressed at all
times. The Hawking flux therefore arises entirely from the first term
in \eqn{2eq44}, in the standard case.]
\end{thebibliography}
\end{document}